\documentclass{article}

\usepackage{amsmath}
\usepackage{amsfonts}
\usepackage{graphicx}
\usepackage{color}
\usepackage{subfigure}

\newcommand{\dd}{{\mathrm d}}

\newcommand{\ee}{{\mathrm e}}
\newcommand{\ii}{{\mathrm i}}

\newcommand{\bn}{{\boldsymbol{n}}}

\newcommand{\bx}{{\boldsymbol{x}}}

\newcommand{\bX}{{\boldsymbol{X}}}

\newcommand{\fh}{{\hat{f}}}

\newcommand{\PD}[2]{\frac{\partial {#1}}{\partial {#2}}}
\newcommand{\PDD}[2]{\frac{\partial^2 {#1}}{\partial {#2}^2}}
\newcommand{\Pd}[2]{{\partial {#1}}/{\partial {#2}}}
\newcommand{\OD}[2]{\frac{\dd {#1}}{\dd {#2}}}
\newcommand{\Od}[2]{{\dd {#1}}/{\dd {#2}}}

\title{Mathematical Modelling of Tyndall Star Initiation}
\author{Andrew A.~Lacey, Matthew G.~Hennessy, Peter Harvey, and Richard
  F.~Katz}

\begin{document}
\maketitle

\begin{abstract}
The superheating that usually occurs when a solid is melted by
volumetric heating can produce irregular solid-liquid interfaces.
Such interfaces can be visualised in ice, where they are sometimes
known as Tyndall stars. This paper describes some of the
experimental observations of Tyndall stars and a
mathematical model for the early stages of their evolution.
The modelling is complicated by the strong crystalline
anisotropy, which results in an anisotropic kinetic undercooling
at the interface; it leads to an interesting class of
free boundary problems that treat the melt region as 
infinitesimally thin.
\end{abstract}


\section{Introduction}                 \label{sec:Intro}

\begin{figure}
  \centering
  \includegraphics[width=0.4\textwidth]{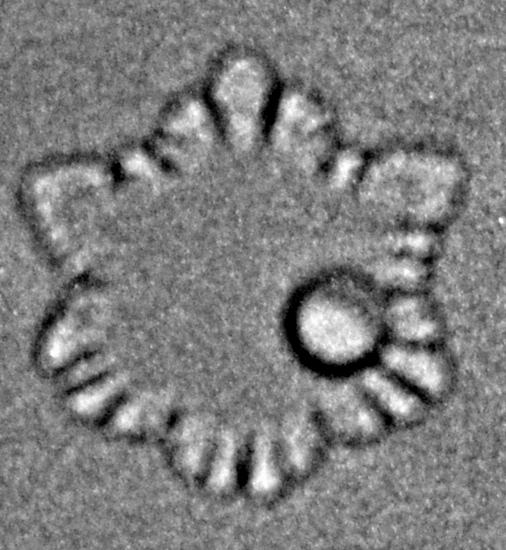}
  \caption{An example of a Tyndall star that has been created by irradiating a 
    pure crystal of
    ice with light from an overhead projector.  
    The bright circle within the star is a vapour
    bubble that emerges due to the density difference between water and ice.  
    The viewing
    plane corresponds to the basal plane of the melting
    ice crystal with the $c$ axis pointing
    orthogonally into and out of the page. This image was created by the authors at the FoaLab in Oxford; for additional details, see Harvey \cite{harvey2013}.}
  \label{fig:exp}
\end{figure}

When a single crystal of pure, transparent ice is irradiated, the
partial absorption of transmitted radiation volumetrically heats the crystal,
leading to
internal melting and the formation of small volumes of liquid. 
Remarkably, these volumes of water often take on shapes that resemble
six-fold symmetric flowers, stars, or snowflakes, as first documented 
by Tyndall \cite{Tyndall}. The internal melt figures that Tyndall
observed now bear his name and are often referred to as Tyndall stars,
Tyndall figures, or liquid snowflakes. An examples of such can be found in
Fig.~\ref{fig:exp}.

\

Tyndall stars are predominantly found in very pure crystals of irradiated 
ice. The
lack of impurities and microscopic defects in such crystals limits the 
onset of liquid nuclei and
prevents the ice from simply melting away as it continually absorbs
radiation. Instead, the ice becomes superheated, whereby its
temperature exceeds the equilibrium melting temperature. It is this
superheating that, through an interfacial
instability, is suspected of giving rise to the complex morphologies
that are characteristic of Tyndall stars. The six-fold symmetry that is 
apparent in Fig.~\ref{fig:exp} is inherited from the
anisotropy of the ice crystal, which will be discussed in detail below. 

\

From a scientific viewpoint, Tyndall stars offer a convenient route for
studying the dynamics of phase change and moving interfaces
because both the solid and liquid phases are transparent. Thus, 
in principle, these phases can be observed in real time with visible light. 
Understanding of Tyndall stars may also have industrial implications in,
for example, resistance welding, 
whereby a metal is volumetrically heated by passing an electrical current
through it \cite{A, LT}. This leads to a superheated solid and the formation
of small inclusions of liquid metal. Due to the opacity of the metal, these
inclusions cannot be seen in real time and are often detected after the welding
operation is over. 

\

The evolution of Tyndall stars has been studied experimentally by
Nakaya \cite{nakaya1956}, who found that the melts begin as cylindrical
discs of water with thicknesses that are much smaller than their radii. 
This thin aspect ratio is maintained during the evolution of a Tyndall
star, with growth in the radial direction being 
much faster than in the axial direction. As the cylindrical disc increases in
size,
the circular interface can become unstable, leading to the emergence of 
a high-wavenumber fingering pattern. In cases where the radiation intensity was
sufficiently high, further growth of the instability resulted 
in the formation six large symmetric dendrites. 
In addition, Nakaya reported that the Tyndall stars in a given ice crystal
always have the same orientation.  
Further experiments by Takeya \cite{T}
were able to provide quantitative data for the radial and axial growth 
of Tyndall stars. Over the duration of a couple of minutes, the radius
increased to roughly 1.5 mm while the thickness grew linearly with time
to about 0.3 mm. In some cases, however, the axial growth of the melt was
only temporary and eventually it stopped altogether. Interestingly, Takeya
reported that an interfacial instability only occurs when the axial growth
persists; in cases where the axial growth terminates, the melt remains
cylindrical.\footnote{The axial growth ceased for cases of low superheating
with there being sufficient heat to melt only a small part
of the ice. There could be only limited scope for instability
in such situations.}
This observation is perhaps linked to those made by
Mae \cite{mae1975}, who found that Tyndall stars retain their
initial cylindrical shape unless they grow beyond a critical thickness of 10
$\mu$m. Experimental \cite{SF} and theoretical \cite{YSF} studies of 
solidification in supercooled liquids, a situation that closely parallels
melting into a superheated solid, have also shown that a critical thickness
must be surpassed in order for a morphological instability to occur at the
solid-liquid interface. 

\

\begin{figure}
  \centering
  \includegraphics[width=0.5\textwidth]{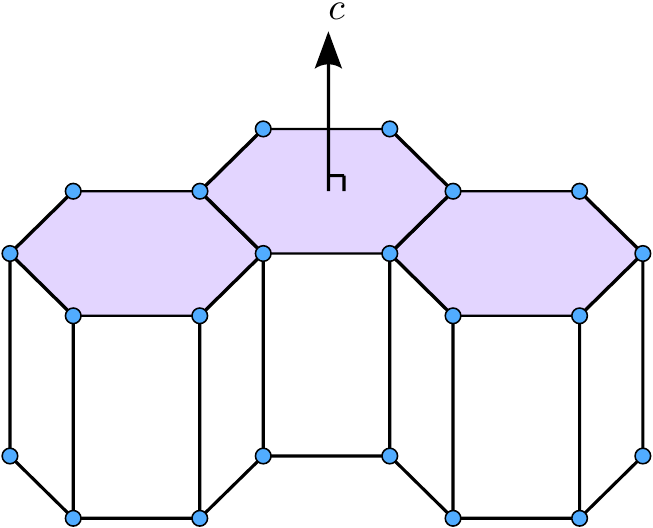}
  \caption{A schematic diagram of an ice crystal, which is composed of arrays of hexagonal
	prisms.  Shaded hexagonal faces form the molecularly smooth basal 
	planes of the ice crystal and the unshaded rectangular faces correspond to molecularly
	rough prism planes.  The $c$ axis of the ice crystal
	is orthogonal to the basal planes.  The 
	shaded circles give the approximate positions of oxygen atoms.  
	The rate of melting is much higher at prism planes than basal planes, 
	resulting in Tyndall stars that are relatively thin in directions along the 
	$c$ axis.}
  \label{fig:crystal}
\end{figure}

The anisotropic growth of a Tyndall star is closely related to the
geometric configuration of the melting ice crystal. Roughly speaking,
the crystalline structure of ice can be imagined as a collection of
adjacent hexagonal prisms; see Fig.~\ref{fig:crystal}.
The hexagonal faces of the prisms form the so-called
basal planes of the crystal and the direction that is normal to these planes
defines the $c$ axis. The radial growth of Tyndall stars occurs within the
basal planes while the axial growth is aligned with the $c$ axis, therefore
giving different Tyndall stars the same orientation within an ice crystal. 
The molecularly smooth basal planes melt at a much slower rate than the
molecularly rough prism planes. As discussed in the context of solidification
\cite{CMW},
the accretion of material normal to a molecularly smooth surface, such as 
a basal plane, occurs via an energetically activated process, whereas 
there is no nucleation barrier at a molecularly rough surface. 
The fast-melting prism planes dominate the shape of the Tyndall figure
\cite{M} and are responsible for the disparity between its axial and
radial dimensions.

\

The mathematical study of problems involving
phase change is now a classical subject for which there is extensive 
literature. Davis \cite{D} gives a comprehensive treatment of the mathematical
theory of solidification, starting from the classical Stefan problem. 
Hu \& Argyropoulos \cite{HA} provide an overview of modelling
and computational techniques that are relevant to solidification and melting
problems. The fluid mechanics of solidification are reviewed in detail by
Huppert \cite{H}. Coriell {\it et al.}~\cite{CMS,CMSB} examine the occurrence of
multiple similarity solutions, as well as their selection mechanisms, in
models of solidification and melting. The application of phase-field models
to solidification problems has been discussed by Boettinger {\it et al.}
\cite{B}.

\

Mathematical models of phase change that account
for the anisotropic nature of the solid have been largely confined to
the case of solidification and crystallisation. 
Wettlaufer {\it et al.}~\cite{CMW, MKKW, TW, W} 
examined two-dimensional crystallisation within
the basal plane by considering an interfacial velocity that depends
on the angle between the free boundary and a certain
fixed direction. A suitable angular dependence was found to give rise 
to the six-fold symmetry that is characteristic of snowflakes. It is 
important to emphasise here that in the studies of Wettlaufer {\it et al.},
it is assumed that growth of the crystal is in the geometric limit, whereby the
interface velocity is only a function of the shape and position of the
interface. In particular, the velocity of the interface
does not depend on field variables that are affected by its motion. 
This is in contrast to non-geometric growth models,
which account for long-range diffusion of field variables and their
coupling to the interfacial velocity. 
In geometric models, the
crystalline anisotropy enters directly through the interface velocity.
However, in non-geometric models, anisotropy enters through physical parameters 
related to the interface, such as surface energy or the coefficient of
kinetic undercooling, 
the latter of which connects the temperature and velocity at the interface. 
Anisotropic solidification outside of the geometric limit has been 
investigated by a number of authors. 
Uehara \& Sekerka \cite{US} studied the formation of facets due to
strong anisotropy in the kinetic coefficient using a phase-field model. 
Particular attention
was paid to determining the relationship between the shape of the emerging 
crystal and the mathematical properties of the anisotropic kinetic coefficient.
Yokoyama \& Kuroda \cite{YK} employed the boundary-element method to study 
the hexagonal morphologies of snow crystals predicted by a model with an
anisotropic kinetic coefficient. 
Yokoyama \& Sekerka \cite{YS} explored the combined effects of anisotropic
kinetic undercooling and surface energy. Using numerical and asymptotic methods,
they investigated the suppression of corner formation between adjacent facets.

\

Considerable attention has focused for many years on the stability of
the free boundary in phase-change models. Linear stability analyses
of models which treat the phase interface as infinitesimally thin, such as
in the pioneering study by Mullins \& Sekerka \cite{MS} or in
Hele-Shaw and Muskat problems, 
indicate that a morphological instability can arise
when a melting boundary is driven by heat flow from a
superheated solid region \cite{LS}. In fact, without a regularising mechanism
such as surface energy or kinetic undercooling,
the system is severely
unstable and the model becomes ill posed in the sense
that disturbances with arbitrarily large wavenumbers
will grow arbitrarily fast in time. 
Such ill-posedness can also be avoided by replacing
the sharp, infinitesimally thin interface with a diffuse mushy region 
consisting of two co-existing phases \cite{A, LS}. 
The theory of mushy regions in volumetrically
heated solids has been developed by Lacey {\it et al.}~\cite{LH1, LH2, LT}, who
treated the mush as a collection of small liquid inclusions that grow within
the solid. 
In these papers, the growth of the inclusions is modelled using 
classical Stefan problems that account for surface-energy effects 
and interfacial
curvature, kinetic undercooling, and/or composition in the case of alloys. 
The main purpose of those studies was to use homogenisation to build an
averaged model for the mushy region.

\

A sharp-interface model of Tyndall stars has been formulated and studied
by Hennessy \cite{hennessy2010}. The focus here was on two-dimensional evolution
within the basal plane. The morphology of the solid-liquid interface was 
studied using a combination of linear stability theory and numerical 
simulations. Growth along the $c$ axis was not considered and thus it was
not possible to explore how this may influence the stability of the ice-water
interface. 

\

In this paper, we consider the three-dimensional evolution of a Tyndall
star or, perhaps more accurately, a Tyndall figure, as we mostly
discuss the earlier growth rather than the
later, star-like stage. Particular attention is paid to capturing the
anisotropic growth along the radial and axial directions. Our description
of the problem is based on the classical Stefan model but the inclusion of
volumetric heating and anisotropic kinetic undercooling makes it non-standard.
An asymptotic analysis that exploits the axial and radial length-scale 
separation is used to reduce the three-dimensional 
problem to a co-dimension-2 free boundary problem whereby the melt is
collapsed into a planar surface with infinitesimal thickness. 
A local stability analysis
of the reduced model is carried out as a first step towards
the study of the onset of fingering
patterns at the ice-water interface. An attempt is made to compare our
theoretical results to the experimental observations of Takeya \cite{T}; 
however, this is not straightforward due to a lack of knowledge
of key quantities controlling the anisotropic growth. We then propose
future experiments that could produce novel quantitative insights
into the growth kinetics. 

\

In the next section, we present a mathematical model for
a growing Tyndall figure based on laboratory experiments. 
In  Sec.~\ref{sec:Theory}, we carry out an asymptotic 
analysis of this model that captures the anisotropic growth
of the melt and investigates the stability of the ice-water
interface. We discuss our results and conclude the paper in 
Sec.~\ref{sec:Discuss}.


\section{Mathematical Model}                 \label{sec:Theory}


\subsection{The Physical Problem}             \label{subsec:physics}

\begin{figure}
  \centering
  \includegraphics[width=0.5\textwidth]{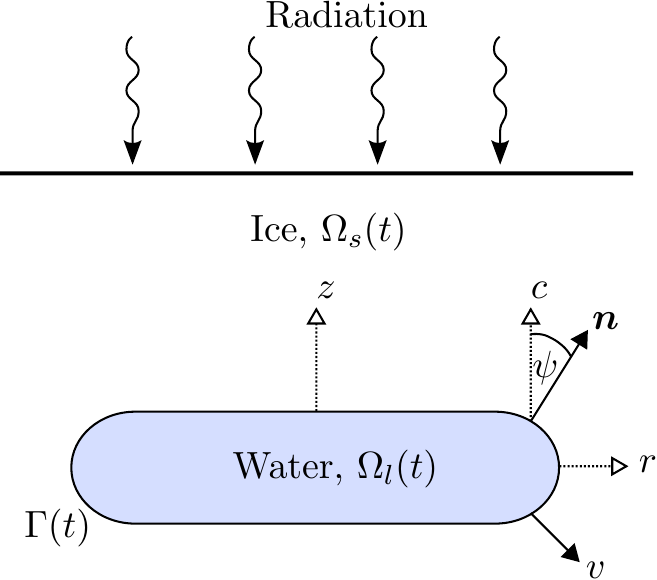}
  \caption{We study the growth of a Tyndall figure (depicted by the
    shaded region) in superheated
    irradiated ice. 
    We use $\Omega_l$ and $\Omega_s$ to denote
    regions of space occupied by liquid water and solid ice,
    respectively.  Here, $t$ represents time.  The ice-water interface
    is denoted by $\Gamma(t)$ and has a normal vector $\bn$ and
    normal component of velocity $v$.  The $z$ axis is parallel to the
    $c$ axis of the ice crystal and $r = (x^2 + y^2)^{1/2}$ is a radial
    coordinate that lies within the basal plane.  The angle between
    the $c$ axis and the normal vector $\bn$ is given by $\psi$.}
  \label{Fig:iceandwater}
\end{figure}

We suppose that a single crystal of ice held at its melting temperature
is illuminated at time $t = 0$.  The direction of the 
incident light is taken to be parallel to the c-axis of the crystal; see
Fig.~\ref{Fig:iceandwater}.
We assume that a rapid nucleation process occurs within the ice
upon exposure to light, leading to the creation of a 
single spherical melt figure. 
Continued absorption of radiation by both the ice and the water
will drive the melting at the interface,
which we aim to describe mathematically.
Our model of this physical scenario is based on equations governing the
temperatures in the liquid and solid phases, taking into account
thermal diffusion and volumetric heat generation due to 
absorption of radiation.
The solid-liquid interface is assumed to be sharp and, therefore, 
we impose appropriate boundary conditions on it.

\

The field equation for the temperature $T_j$ of phase $j$ is 
given by
\begin{equation}
\rho_j c_{j}\PD{T_j}{t}= k_j \nabla^2 T_j + q_j, \quad
\bx \in \Omega_j(t)
\label{eq:field:dim}
\end{equation}
where $t$ is time, position is $\bx = (x,y,z)$, and $\Omega_j(t)$ 
is the region of space occupied by phase $j$. We let $j = l$ 
and $j = s$ for the liquid water and solid ice phases, respectively.
We assume
that the $z$ axis and the $(x,y)$ plane are aligned with the c axis
and basal planes of the ice crystal, respectively.
The values of the material constants, namely the densities,
$\rho_j$, specific heat capacities, $c_{pj}$, and thermal conductivities,
$k_j$, differ between the two phases. Although
the difference in density between the phases is significant enough
to give rise to a vapour bubble inside the Tyndall
figure, as shown in Fig.~\ref{fig:exp}, their relative
difference is small and we take the densities of the 
two phases to be the same and equal to $\rho$, that is,
$\rho_l = \rho_s = \rho$.
The rates of volumetric
heating, $q_j$, are given by the product of an absorption coefficient,
$\mu_j$, and the local intensity of incident light upon the medium, $I$.
With a sufficiently small piece of ice
(or absorption coefficient), $I$
can be regarded as constant, making $q_j$
constant in each phase. We shall generally assume that
the initial temperatures coincide with
the equilibrium melting temperature $T_0$
at $t=0$, with a spherical Tyndall figure of radius $a$
nucleating at the same instant. However, if significant
body heating occurs before nucleation, the initial
temperatures will be much greater than $T_0$. This 
situation is discussed in Appendix \ref{app:nucleation}.

\

At the evolving interface $\Gamma = \Gamma(t)$ 
between ice and water, we have the usual Stefan condition
\begin{equation}
L\rho v = \left[ k_j \PD{T_j}{n} \right]_l^s , \quad \bx \in \Gamma(t),
\label{eq:Stefan:dim}
\end{equation}
where $L$ is the latent heat of fusion, assumed
constant; $v$ is the normal velocity, measured
towards the ice; $\Pd{}n$ is the normal derivative,
again in the direction into the ice;
and $[ \cdot ]_l^s$ denotes the change in a quantity across
the interface, going
from liquid water to solid ice, see Fig.~\ref{Fig:iceandwater}.

\

We also assume that the normal velocity of the interface
is proportional to the local amount of superheating \cite{D}. 
To account for the
different melting rates of the basal and prism planes, we take the constant of
proportionality to be a function of the orientation of the interface. 
Thus, we impose a kinetic condition, 
equivalent to anisotropic kinetic undercooling in solidification 
\cite{US, YK, YS}, given by
\begin{align}
  v = K f(\psi)(T_I - T_0), \quad \bx \in \Gamma(t),
\label{eq:kinetic:dim}
\end{align}
where $T_I$ is the temperature at the interface,
\begin{align} 
  T_I = T_s = T_l, \quad \bx \in \Gamma(t);
\end{align}
$K$ is a constant; $f$ is a dimensionless function, which we refer to as the
anisotropy function; and $\psi$ is the angle between normal vector at
the free surface and the
$c$ axis, as measured relative to the positive $x$ axis, 
see Fig.~\ref{Fig:iceandwater}. 
Contributions to \eqref{eq:kinetic:dim} from the surface energy are not
included, which we justify by assuming that after the rapid nucleation phase,
the radius of the melt is much larger than the capillary length given by
$l_\mathrm{cap} = (\gamma / \rho L) (T_0 / \Delta T)$, where $\gamma$ is the
surface energy of an ice-water interface and $\Delta T$ is the
local amount of superheating. Takeya \cite{T}
measured superheatings on the order of 0.1 K in his experiments
that use photographic bulbs as the light source, 
giving a capillary length of 270~nm; thus, neglecting surface energy
seems reasonable given that Tyndall figures typically
have length scales on the order of
hundreds of microns up to millimetres.
 An important consequence of neglecting surface energy 
is that our model will not capture the evolution of the
melt into a Wulff shape, which is the equilibrium shape arising from the
minimisation of surface energy under constant-volume conditions \cite{D}. 
However, based on the phase-field simulations by Uehara \& Sekera \cite{US},
we might expect the melt to grow into its ``kinetic Wulff shape'', which, in essence,
describes the asymptotic shape that the interface would approach 
if it were to evolve solely due to anisotropic undercooling under isothermal
conditions, so that the normal velocity depends only upon
the orientation of the interface \cite{US,YK, YS}
(also see, below, Sub-sec~\ref{subsec:early} and
Sub-sec~\ref{subsec:orderone}).


\subsection{The Anisotropy Function}  

\begin{figure}
  \centering
  \includegraphics[width=0.8\textwidth]{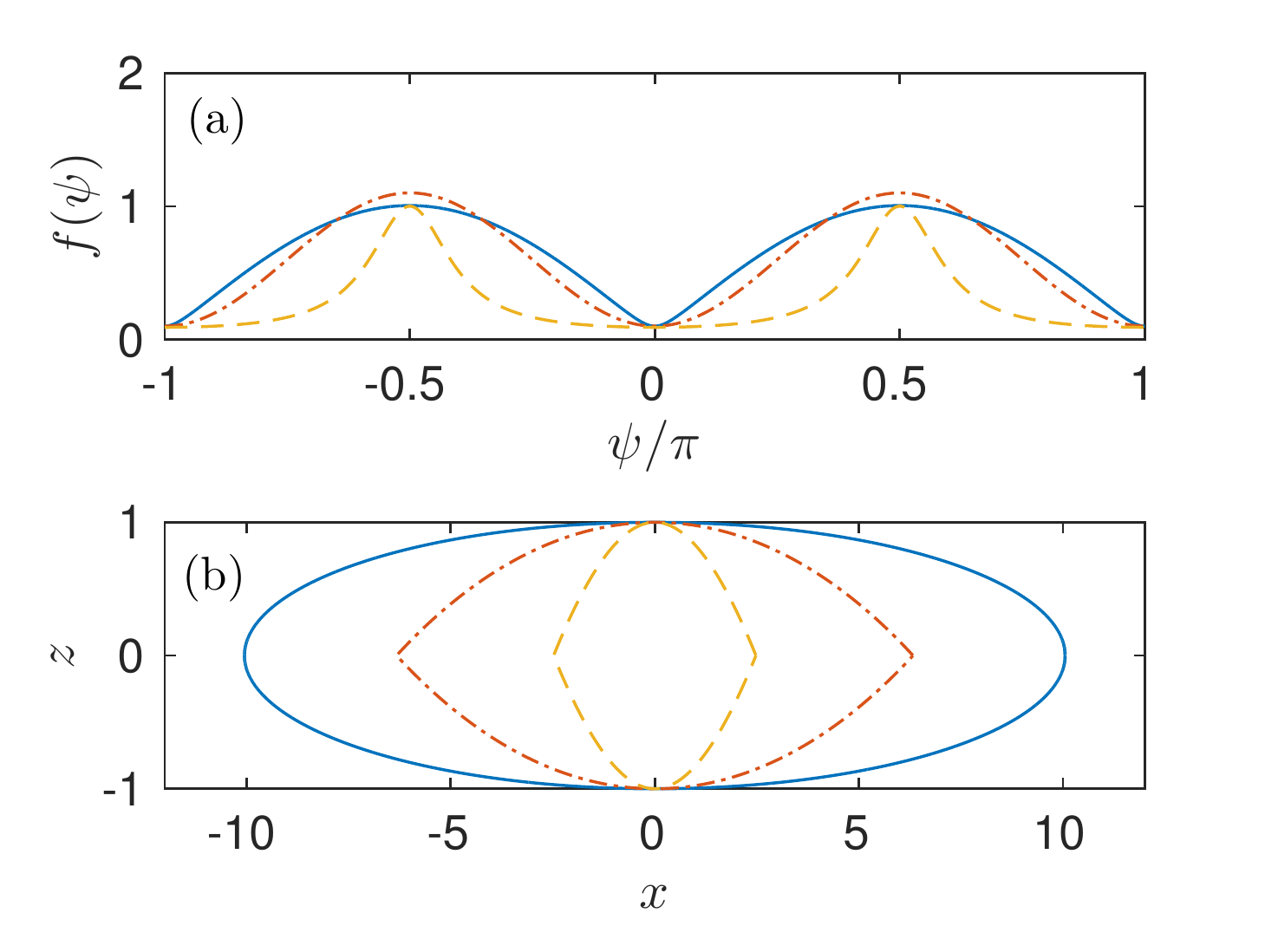}
  \caption{Top (a): we consider three different anisotropy 
    functions $f$ that characterise the dependence
    of the melting rate on the orientation of the solid-liquid 
    interface: $f(\psi) = (\epsilon^2 + \sin^2 \psi)^{1/2}$
    (solid), $f(\psi) = \epsilon + \sin^2 \psi$ (dash-dotted), 
    and $f(\psi) = \epsilon / (1 + \epsilon - \sin^2 \psi)$
    (dashed).  Here, $\psi$ measures the angle between the $c$ axis and the 
    vector normal to the interface;
    see Fig.~\ref{Fig:iceandwater}.  Bottom (b): the corresponding kinetic 
    Wulff shapes associated with the
    three anisotropy functions $f$, which represent the long-term shape the 
    melt would acquire under 
    isothermal conditions and growth due purely to anisotropic kinetic 
    undercooling.  See text for
    further details.  In both panels we have set $\epsilon = 0.1$.}
  \label{fig:kin_wulff}
\end{figure}

The anisotropy function $f$ is used to model the orientation dependence of the
interfacial velocity arising from the crystalline structure of the ice. 
We assume that the value of $f$ is close to one
when the velocity is parallel to the prism planes of the ice crystal
and small when the velocity is parallel to the basal planes.
Mathematically, this corresponds to
$f \sim 1$ when $\psi = \pm \pi/2$, and $f \sim \epsilon \ll 1$ 
when $\psi = 0, \pm\pi$, respectively. 
In physical terms, the parameter $\epsilon$ can be thought of as
the ratio of the melting velocity of basal planes
to prism planes for a fixed superheating $T_I - T_0 > 0$. 
Experimentally determining a functional form for $f$ is possible by
measuring the kinetic Wulff shape. However, acquiring the kinetic
Wulff shape is difficult in practice and,
consequently, there is often uncertainty in the form of $f$.  
Therefore, our analysis will rely on phenomenological expressions for the
anisotropy function. More specifically, we will consider in detail
the function
\begin{align}
  f(\psi) = (\epsilon^2 + \sin^2\psi)^{1/2},
  \label{eqn:f}
\end{align}
which is expected to produce smooth interfaces based on 
its corresponding kinetic Wulff shape. In two-dimensions,
the kinetic Wulff shape is determined by the convex region
containing the origin traced out by the parametric curves
\begin{subequations}
\label{eqn:kin_wulff}
\begin{align}
  x &= f'(\psi) \cos \psi +  f(\psi) \sin \psi, \\
  z &= f'(\psi) \sin \psi - f(\psi) \cos \psi.
\end{align}
\end{subequations}
The anisotropy function \eqref{eqn:f} is shown along with its corresponding
Wulff shape in Fig.~\ref{fig:kin_wulff}.
Additionally, we will present the key results that are obtained when
\begin{align}
  f(\psi) = \epsilon + \sin^2 \psi,
  \label{eqn:f2}
\end{align}
and
\begin{align}
  f(\psi) = \frac{\epsilon}{1 + \epsilon - \sin^2 \psi}.
  \label{eqn:f3}
\end{align}
Since $\sin^2 \psi$ can be written in terms of $\sin 2\psi$, the anisotropy
function \eqref{eqn:f2} is similar to many of those found in the literature
\cite{US}. The anisotropy function in \eqref{eqn:f3} has sharp maxima at
$\psi = \pm \pi / 2$ (see Fig.~\ref{fig:kin_wulff} (a)), making it comparable to
theoretical expressions for $f$ that have been derived from models of
surface diffusion \cite{BCF,YK}. Figure \ref{fig:kin_wulff} shows that the
anisotropy functions \eqref{eqn:f2} and \eqref{eqn:f3} lead to the formation
of corners in the kinetic Wulff shape. Surface energy is likely to become
important on these small scales and may lead to a smoothing of the corners.
Capturing such dynamics is beyond the scope of our current model, however. 

\subsection{Parameter Values}

The configuration that we study here is based on experiments 
involving ice-water systems carried out in Oxford. Light from
an overhead projector was used to irradiate a pure ice crystal. 
A list of parameter values corresponding to these experiments is given
in Table \ref{tab:params}.  Although light from the overhead projector
will have a broad spectrum, ice and water are particularly strong 
absorbers of infra-red radiation. Therefore, the absorption coefficients
in Table \ref{tab:params} are based on monochromatic infra-red radiation with 
a wavelength of 980~nm.  The intensity of radiation has been calculated from
the bulb power and distance to the sample by assuming spherical emission;
the complete details can be found in Hennessy \cite{hennessy2010}.

\begin{table}
\centering
\caption{Parameter values for ice-water systems heated by light
	from an overhead projector.  These are based on experiments carried
	out in Oxford.  The absorption coefficients are for monochromatic
	infra-red radiation with a wavelength of 980~nm.  The intensity of
	radiation is estimated from the power of the bulb and distance to the
	sample, further details are given in the text.}
\label{tab:params}
\begin{tabular}{cc}
\hline
$\rho$ & 1000 kg/m$^3$ \\
$c_{ps}$ & 2050 J/(kg K) \\
$c_{pl}$ & 4181 J/(kg K) \\
$k_s$ & 2 W/(m K) \\
$k_l$ & 0.6 W/(m K) \\
$L$ & $3.33\times 10^5$ J/kg \\
$\mu_s$ & 15.3 1/m \\
$\mu_l$ & 43.6 1/m \\
$I_0$ & 300 W/m$^2$ \\
$\gamma$ & 0.033 J/m$^2$ \\
$T_0$ & 273 K \\ \hline
 \end{tabular}
\end{table}

\

Determining values for the parameters $K$ and $\epsilon$ is a challenging
experimental task. 
Using arguments from statistical mechanics, it is possible to write
the velocities of the planes \cite{WJE}, as well the coefficient 
$K$ in \eqref{eq:kinetic:dim} \cite{FMM}, in terms of elementary quantities such as
molecular distance and activation energy. However, these expressions
introduce additional unknown parameters into the model, making them of
little practical use. 
The combined uncertainty in the values for $K$ and $\epsilon$, as well as in
the functional form of the anisotropy function
$f$, will make carrying out a quantitative comparison of our results with
experimental data difficult. That being said, qualitative comparisons are still
possible, and the analysis can be used as a tool for ruling out anisotropy
functions. 


\subsection{Non-dimensionalisation}             \label{subsec:nondim}

The model is non-dimensionalised by introducing
suitable scales for time, distance, and temperature.
The time variable $t$ is written in terms of the
time scale of thermal diffusion in ice, $\ell^2 / \kappa_s$,
where $\kappa_s = k_s / (\rho c_{ps})$ is the thermal
diffusivity of ice and $\ell$ is a characteristic length scale
defined below. 
The temperature scale is set by the
amount of superheating in the ice caused by 
volumetric heating, giving $\Delta T = q_s \ell^2 / k_s$. 
Finally, the length scale $\ell$ is chosen to balance 
terms in the kinetic condition \eqref{eq:kinetic:dim},
implying that significant growth parallel to the basal planes
occurs on $O(1)$ (dimensionless) time scales. This gives
$\ell^3 = K k_s^2 / (q_s \rho c_{ps})$.
Using these scales, we write
$t = (\ell^2 / \kappa_s) \tau$, $\bx = \ell \bX$, 
and $T_j = T_0 + (\Delta T) \theta_j$.
The non-dimensional field equations can be written as
\begin{subequations}
  \label{eq:nondim}
\begin{alignat}{2}
  \PD{\theta_s}{\tau} &= \nabla^2 \theta_s + 1, 
  &\quad \bX &\in \Omega_s(\tau), \\
  \hat{c}_p\PD{\theta_l}{\tau} &= \hat{k} \nabla^2 \theta_l + \hat{q}, 
  & \quad \bX &\in \Omega_l(\tau),
\end{alignat}
where $\hat{c}_{p} = c_{pl}/c_{ps}$ and $\hat{k} = k_l / k_s$ 
are ratios of specific heat capacities and thermal conductivities,
respectively. The ratio of volumetric heating, $\hat{q} = q_l / q_s$,
can be written in terms of the absorption coefficients via
$\hat{q} = \mu_l / \mu_s$. Initial conditions for the
temperatures are given by $\theta_s = \theta_l = 0$ at $\tau = 0$.

At the ice-water interface, the
Stefan and kinetic conditions, along with the continuity of temperature,
are given by
\begin{alignat}{2}
  v &= \beta^{-1} \left(\PD{\theta_s}{n} - \hat{k}\PD{\theta_l}{n}\right),
  &\quad
  \bX &\in \Gamma(\tau), \\
  v &= \theta_I f(\psi), &\quad \bX &\in \Gamma(\tau),\\
  \theta_I &= \theta_s = \theta_l,, &\quad \bX &\in \Gamma(\tau),
  \label{eqn:kinetic:nondim}
\end{alignat}
respectively,
where $\beta = L / (c_{ps} \Delta T)$ is the Stefan number.
The initial ice-water
interface is taken to be the sphere with dimensionless radius
$\alpha = a / \ell$ given by $|\bX| = \alpha$.

\

Far from a growing liquid inclusion,
$\Pd{\theta_s}{\tau} \sim 1$, so that we have
\begin{equation}
 \theta_s \sim \tau, \quad |\bX| \to \infty.
\label{eq:infinite:nondim}
\end{equation}
Note that \eqref{eq:infinite:nondim} requires that the Tyndall
figure and associated length scales be small compared
with the region subject to the body heating.
\end{subequations}

\

Using the parameter values in Table \ref{tab:params}, we 
find that $\hat{k} \simeq 0.3$, $\hat{c}_p \simeq 2$, 
and $\hat{q} \simeq 3$, all of which can be treated
as $O(1)$ in size. Due to uncertainty in the value of
the parameter $K$, it is difficult to estimate the
length scale $\ell$, the characteristic 
temperature rise $\Delta T$, and the Stefan number $\beta$.
Using instead the measured value of $\Delta T \sim 0.1$ K from Takeya 
\cite{T}, the Stefan number is given by $\beta \sim 10^3$.
The length scale can be estimated from 
$\ell = (\Delta T k_s / q_s)^{1/2} \sim 6.7$ mm
and the time scale from $\ell^2 / \kappa_s \sim 46$ s,
which seem slightly large but reasonable. 

\

The proceeding analysis will focus on the
distinguished limit whereby
$\epsilon = O(\beta^{-1})$.
This regime is considered so that we can
examine the interplay of the kinetic anisotropic effects;
whether or not this balance occurs in practice depends
upon the size of the rate of the volumetric heating.
Thus, we write $\beta^{-1} =
b \epsilon$ where $b = O(1)$. Furthermore, it will
be assumed that the (dimensionless) radius of the 
initial melt, $\alpha$, satisfies $\alpha \ll \epsilon$. 
In dimensional terms, this inequality means that the
initial radius should be less than one micron, which
is close to the limit where surface-energy effects become
important. This upper bound on the initial size
of the radius, along with the anisotropic
kinetic condition \eqref{eqn:kinetic:nondim}, 
ensures that the spherical Tyndall figure will
first grow into a thin disc of melt with radius that is 
much greater than its thickness, which is
consistent with experimental observations \cite{T}.


\section{Analysis}

The analysis begins in Section \ref{subsec:early}
with an examination of the small-time
behaviour for $\tau = O(\alpha^{1/2})$. In dimensional terms,
the small-time regime corresponds to times given by
$t \sim (a/\ell)^{1/2}(\ell^2 / \kappa_s)$. Taking the
dimensional radius of the initial melt to be of the order
of one micron, we find that $t \sim 0.5$ seconds.
In this regime,  the volumetric heating and
the kinetic condition drive the melt into a thin shape
with dimensions along the $c$ axis that are much smaller than
those in the basal plane. In Section \ref{subsec:orderone},
we consider the dynamics when $\tau = O(1)$, corresponding to
$t \sim 50$ seconds. By exploiting
the separation of length scales that arises from the initial
growth, a simplified model can be derived. Using this model,
the linear stability of the ice-water interface is examined in
Section \ref{subsec:stability}. Our analysis will first focus on
the dynamics that occur when the anisotropy function \eqref{eqn:f}
is used. We will then consider additional anisotropy functions
in Section \ref{subsec:otheranisot}.


\subsection{Early Time}             \label{subsec:early}

The analysis of the early-time behaviour proceeds by
letting $\tau = \alpha^{1/2} \bar{\tau}$, 
$\theta_j = \alpha^{1/2} \bar{\theta}_j$,
where $\alpha \ll \epsilon \ll 1$. We then consider
the temperature field near and away from the melt,
and connect the solutions in the two regions using 
asymptotic matching. 

\

In the region of solid away from the melt, i.e.,
for $\bX \sim O(1)$, the leading-order problem in $\alpha$ 
is straightforward to solve and it gives
$\bar{\theta}_s(\bX, \bar{t}) = \bar{\tau}$. To resolve
the temperatures near the melt, we let
$\bX = \alpha \bar{\bX}$.
The leading-order problem in
$\alpha$ in this inner region is given by
\begin{alignat}{2}
  \nabla^2 \bar{\theta}_s &= 0, \quad \bar{\bX} \in \bar{\Omega}_s(\bar{\tau}),
  \\
  \nabla^2 \bar{\theta}_l &= 0, \quad \bar{\bX} \in \bar{\Omega}_l(\bar{\tau}),
\end{alignat}
with the following conditions at the solid-liquid interface:
\begin{alignat}{2}
  \PD{\bar{\theta}_l}{n} &= \hat{k} \PD{\bar{\theta}_s}{n}, &\quad
  \bar{\bX} &\in \bar{\Gamma}(\bar{\tau}), \\
  \bar{v} &= \bar{\theta}_I f(\psi), &\quad
  \bar{\bX} &\in \bar{\Gamma}(\bar{\tau}).
\end{alignat}
By asymptotically matching the temperatures in the solid,
we also have that $\bar{\theta}_s \to \bar{\tau}$
as $|\bar{\bX}|
\to \infty$. The solutions for the temperature fields
are given by $\bar{\theta}_l = \bar{\theta}_s \equiv \bar{\tau}$. 
The motion of the interface, therefore, satisfies the equation 
\begin{align}
  \bar{v} = \bar{\tau} f(\psi).
  \label{eqn:kin1}
\end{align}

\

To make further progress, we suppose that the
rescaled positions of the ice-water interface 
are given by the zero level set of a function
$F$, defined by
\begin{align}
  F = \bar{s}(\bar{\bX}(\bar{\tau})) - \bar{\tau}^2 / 2 \equiv 0,
  \label{eqn:F}
\end{align}
where, $\bar{s}$ a function that is to be determined.
The initial shape of the interface is encoded in
the function $\bar{s}$; we require that
$\bar{s}(\bar{\bX}(0)) \equiv 0$ on the sphere
$|\bar{\bX}| = 1$ when $\bar{\tau} = 0$.
We emphasise here that $\bar{s}$ also plays the role of
a time variable; from  \eqref{eqn:F} we see that
$\bar{s} = \bar{\tau}^2 / 2$.
In this formulation, 
the normal velocity at the interface can be
written as $\bar{v} = \bar{\tau} / |\nabla \bar{s}|$ and,
therefore, the kinetic condition \eqref{eqn:kin1}
becomes
\begin{align}
  |\nabla \bar{s}| f(\psi) = 1.
  \label{eqn:kin2}
\end{align}
Closing the problem requires writing the angle $\psi$
in terms of the function $\bar{s}$. For clarity,
we now consider the two-dimensional problem 
by writing $\bar{\bX} = (\bar{X}, 0, \bar{Z})$.
In this case, simple trigonometry shows that
the angle $\psi$ satisfies 
$\sin \psi = \bar{s}_{\bar{X}} / (\bar{s}_{\bar{X}}^2 + \bar{s}_{\bar{Z}}^2)^{1/2}$,
where $\bar{s}_{\bar{X}} = \partial \bar{s} / \partial \bar{X}$
and $\bar{s}_{\bar{Z}} = \partial \bar{s} / \partial \bar{Z}$. By
writing $f(\psi) = \hat{f}(\sin \psi)$, the kinetic equation
\eqref{eqn:kin2} becomes
\begin{align}
  (\bar{s}_{\bar{X}}^2 + \bar{s}_{\bar Z}^2)^{1/2}
  \hat{f}\left(\bar{s}_{\bar{X}}(\bar{s}_{\bar{X}}^2 + \bar{s}_{\bar{Z}}^2)^{-1/2}
    \right) = 1.
\end{align}

\begin{figure}
  \centering
  \includegraphics[width=0.8\textwidth]{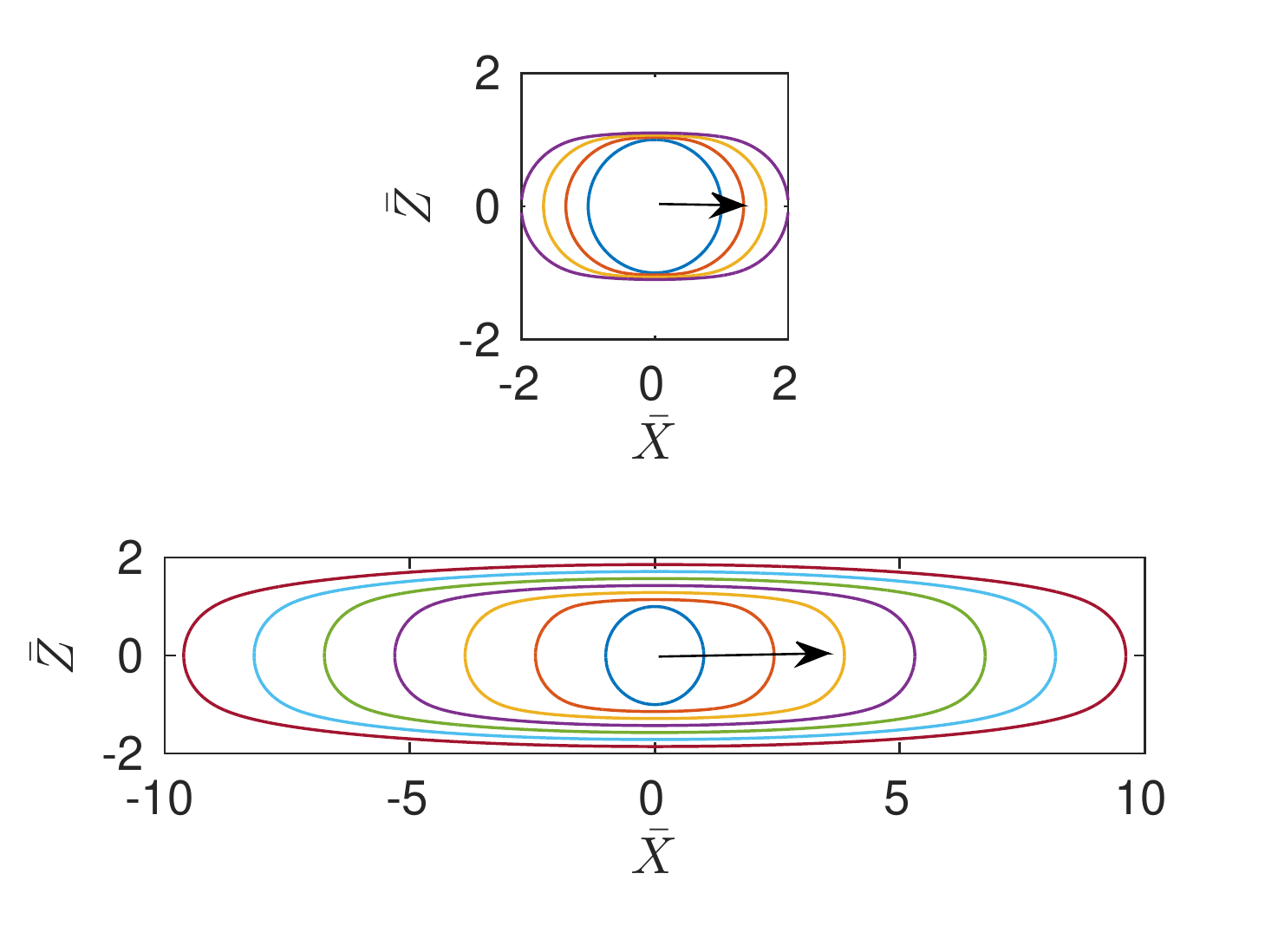}
  \caption{The early-time evolution of a spherical melt figure
    when the anisotropy function is given by \eqref{eqn:f} when
    $\epsilon = 0.1$. These curves are given by the solution in
    \eqref{eqn:char_soln}.
    The arrows indicates the direction of time. The top panels shows the
    solid-liquid interface at equally spaced values of $\bar{s}$ given by	
    $\bar{s} = 0$, $0.33$, $0.66$, and $1$, corresponding to
    rescaled dimensionless times given by
    $\bar{\tau} = (2\bar{s})^{1/2} = 0$, $0.82$, $1.15$, and $1.41$, 
    respectively.
    Similarly, the bottom panel shows the interface for values of $\bar{s}$ 
    given by $\bar{s} = 0$, $1.42$, $2.86$, $4.29$, $5.71$, $7.14$, and $8.57$,
    corresponding to $\bar{\tau} = 0$, $1.69$, $2.39$, $2.93$, $3.38$, $3.78$,
    and $4.14$. 
    The interface remains smooth for all time and evolves into the
    kinetic Wulff shape shown in Fig.~\ref{fig:kin_wulff}.}
  \label{fig:anisotropy_1}
\end{figure}

To see how the melt region evolves, we now focus on
the anisotropy function given by \eqref{eqn:f}. In this
case, the problem for $\bar{s}$ is
\begin{align}
(\bar{s}_{\bar{X}}^2 + \bar{s}_{\bar{Z}}^2)^{1/2} \left[\epsilon^2 + \bar{s}_{\bar{X}}^2 
(\bar{s}_{\bar{X}}^2 + \bar{s}_{\bar{Z}}^2)^{-1}\right]^{1/2} = 1,
\end{align}
subject to the condition
$\bar{s}(\bar{X}_0, \bar{Z}_0) \equiv 0$ on the circle
$\bar{X}_0^2 + \bar{Z}_0^2 = 1$ at time $\bar{\tau} = 0$.
The solution to this problem can be found using Charpit's 
equations, as detailed in Appendix \ref{app:charpit}. 
In essence, Charpit's equations are a generalisation of
the method of characteristics for nonlinear first-order
hyperbolic problems. We proceed by parametrising the
initial data according to
$\bar{\bX}_0(\varphi) = (\bar{X}_0, \bar{Z}_0) 
= (\cos\varphi,\ \sin\varphi)$, 
$\bar{s}(\bar{\bX}_0(\varphi)) \equiv 0$, with 
$\varphi \in [0, 2\pi)$. Upon applying the method,
solution can be written implicitly and parametrically as
\begin{align}
  \bar{X} = \left[1 + \frac{\bar{s}(1 + \epsilon^2)}
    {(\epsilon^2 + \cos^2\varphi)^{1/2}}\right]\cos\varphi, 
  \quad
  \bar{Z} = \left[1 + \frac{\bar{s}\epsilon^2}
    {(\epsilon^2 + \cos^2\varphi)^{1/2}}\right]\sin\varphi, 
  \label{eqn:char_soln}
\end{align}
Thus, for a given value of $\bar{s}$, which can be
written in terms of time via $\bar{s} = \bar{\tau}^2/2$,
these curves trace out the instantaneous positions of
the solid-liquid interface as $\varphi$ is varied from
$0$ to $2\pi$.  
Figure \ref{fig:anisotropy_1} shows the interface profiles predicted by
\eqref{eqn:char_soln} at various times when $\epsilon = 0.1$. 
The initially spherical melt
first grows primarily in the radial direction, keeping its 
thickness in the axial direction constant (top panel). 
By the time the axial growth becomes appreciable, the radius of
the melt has grown a substantial amount, resulting in a 
liquid region with a small aspect ratio.

\

To aid in the physical interpretation of \eqref{eqn:char_soln}, we
revert to the original non-dimensionalisation by writing
$\bar{s} = s / \alpha$, $\bar{\bX} = \bX / \alpha$, 
$\bar{\tau} = \tau / \alpha^{1/2}$ to obtain
\begin{equation}
X = \left[ \alpha + \frac{s(1 + \epsilon^2)}{(\epsilon^2 + \cos^2 \varphi)^{1/2}}
\right] \cos\varphi, \quad
Z = \left[ \alpha + \frac{s \epsilon^2}{(\epsilon^2 + \cos^2 \varphi)^{1/2}}
\right]\sin\varphi,
\label{eq:earlyshape:a1}
\end{equation}
where $s = \tau^2 / 2$.
In the very early stages of development, so that $s$
is of order $\alpha$, then for parts of the interface given by
$|\cos\varphi| \gg \epsilon$,
\begin{align}
Z \sim \alpha \sin\varphi, \quad
X \sim \left( \alpha + \frac s{|\cos\varphi|} \right) \cos\varphi
\sim \pm s + \alpha \cos\varphi,
\end{align}
while for $|\cos\varphi| = O(\epsilon)$, say
$\varphi = \pm \pi/2 \mp \psi$ with $\psi = O(\epsilon)$,
\begin{align}
Z \sim \pm \alpha, \quad
X \sim \pm \left( \alpha \pm \frac s{\epsilon^2 + \psi^2} \right) \psi
\sim \frac{\pm \psi s}{(\epsilon^2 + \psi^2)^{1/2}}.
\end{align}
Thus the interface takes the form,
approximately, of two circular arcs, each of radius $\alpha$
and centred on $(X,Z) = (\pm s,0)$, linked by horizontal lines.

\

In the later stages, $s \gg \alpha/\epsilon$,
\begin{align}
  Z \sim  \frac {\epsilon^2 s}{(\epsilon^2
    + \cos^2 \varphi)^{1/2}} \sin\varphi,
  \quad X \sim  \frac {s(1 + \epsilon^2)}{(\epsilon^2
+ \cos^2 \varphi)^{1/2}} \cos\varphi,
\label{eqn:later}
\end{align}
and
\begin{equation}
\frac{X^2}{1 + \epsilon^2} + \frac{Z^2}{\epsilon^2} \sim \frac {s^2}{\epsilon^2 + \cos^2 \varphi}
[ (1 + \epsilon^2) \cos^2\varphi + \epsilon^2 \sin^2 \varphi]  = s^2 ,
\label{eq:a:appelipse}
\end{equation}
so the interface is then approximately elliptical.
The longer-term 
interface profile, defined by the large-time limit of the small-time model,
and given by \eqref{eqn:later} for this choice of $f$, is, in fact,
equivalent to the corresponding kinetic Wulff shape that can
be computed from \eqref{eqn:kin_wulff}.
Note that the half thickness of the melt, given by the 
maximum value of $Z$, grows in time as 
$Z(\varphi=\pi/2) = \epsilon \tau^2/2$.
The maximum value of $X$, corresponding to the
rim of the melt, grows as 
$X(\varphi = 0) \sim \tau^2 / 2$.
We see that for $s \gg \alpha$, the influence of the initial
interface has been lost.


\subsection{Order-One Time}             \label{subsec:orderone}

We now consider the dynamics that occur on $O(1)$
time scales.
The initial condition in this time regime
takes the form of a matching
requirement, as $\tau \to 0$,
with the fully developed early-time shape given by
\eqref{eq:a:appelipse}. 
The analysis in two and three dimensions is sufficiently
similar for us to proceed directly to problems with
axial symmetry. Thus, we define a radial coordinate
$R = (X^2 + Y^2)^{1/2}$. We also assume the system remains
symmetric about the $Z = 0$ plane and, thus, we only
consider the problem in the upper-half space given by
$Z > 0$.
The position of the solid-liquid interface
is written as $Z = h(R, \tau)$; the corresponding
position of the rim is $R = S(\tau)$ so that
$h(S(\tau), \tau) \equiv 0$. 
The angle $\psi$
appearing in the anisotropy function $f$ 
satisfies
\begin{align}
  \sin \psi = \frac{\Pd{h}{R}}{(1 + (\Pd{h}{R})^{2})^{1/2}}.
\end{align}
From matching into the early-time regime and using 
\eqref{eq:a:appelipse}, we expect that
\begin{align}
h(R) \sim \epsilon(S^2 - R^2)^{1/2}
\label{eqn:ic_h}
\end{align}
as $\tau \sim 0$.

\

\begin{figure}
  \centering
  \includegraphics[width=0.6\textwidth]{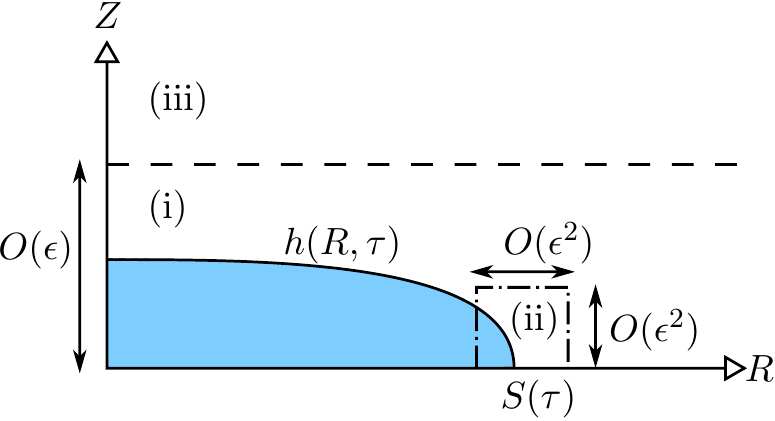}
  \caption{A schematic diagram showing the three asymptotic regions
    in the $\tau = O(1)$ problem. By constructing local solutions
    in regions (i) and (ii), it is possible to derive effective boundary
    conditions that lead to a self-contained problem in region (iii)
    by asymptotic matching.}
  \label{fig:regions}
\end{figure}

In principle, the dynamics in the $O(1)$
time regime can
be studied by solving \eqref{eq:nondim} directly.
However, the thin aspect ratio of the 
melt, with $Z \sim O(\epsilon)$ and
$X,Y \sim O(1)$, motivates seeking a solution
via matched asymptotic expansions,
and this is the approach
we take. There are three distinct regions that need
to be considered: (i) near the melt but away from
the rim, (ii) near the melt and near the rim, and
(iii) away from the melt. A schematic diagram of these
regions is shown in Fig.~\ref{fig:regions}.
Our approach is to obtain
local solutions in regions (i) and (ii) which can then
be used to derive effective boundary conditions for the
problem in region (iii) by asymptotic matching.

\subsubsection{Analysis near the melt and away from the rim}

In region (i) near the melt but away from the rim,
$R \ll S(\tau)$, we rescale the axial coordinate according to
$Z = \epsilon \tilde{Z}$. In addition, the position of the
interface is written as $h(R, \tau) = \epsilon \tilde{h}(R,\tau)$
and the temperatures in this region are denoted by
$\tilde{\Theta}_j$, $j = l,s$. Under this scaling, the anisotropy function
\eqref{eqn:f} can be written as 
$f(\psi) \sim \epsilon [1 + (\Pd{\tilde{h}}{R})^2]^{1/2}$.
The governing equations in this region are given by
\begin{subequations}
\begin{align}
  \epsilon^2 \PD{\tilde{\Theta}_s}{\tau} = \frac{\epsilon^2}{R}
  \frac{\partial}{\partial R}\left(R \PD{\tilde{\Theta}_s}{R}\right)
  + \PDD{\tilde{\Theta}_s}{\tilde{Z}} + \epsilon^2, \quad \tilde{Z} > 
  \tilde{h}(R, \tau), \\
  \epsilon^2 \hat{c}_p\PD{\tilde{\Theta}_l}{\tau} = \frac{\epsilon^2\hat{k}}{R}
  \frac{\partial}{\partial R}\left(R \PD{\tilde{\Theta}_l}{R}\right)
  + \hat{k}\PDD{\tilde{\Theta}_l}{\tilde{Z}} + \epsilon^2 \hat{q}, \quad 
  \tilde{Z} < \tilde{h}(R,\tau).
\end{align}
The boundary conditions on the solid-liquid interface are
\begin{alignat}{2}
  \epsilon b^{-1}\PD{h}{\tau} &= 
  -\hat{k}\left(\PD{\tilde{\Theta}_l}{\tilde{Z}} - \epsilon^2\PD{\tilde{\Theta}_l}{R} 
    \PD{\tilde{h}}{R}\right) 
  + \PD{\tilde{\Theta}_s}{\tilde{Z}} - \epsilon^2\PD{\tilde{\Theta}_s}{R}
  \PD{\tilde{h}}{R}, &\quad \tilde{Z} &= \tilde{h}(R,\tau), \\
  \PD{\tilde{h}}{\tau} &= \tilde{\Theta}_I 
  \Bigg[1 + \left(\PD{\tilde{h}}{R}\right)^2\Bigg]^{1/2}
  \Bigg[1 + \epsilon^2 \left(\PD{\tilde{h}}{R}\right)^2\Bigg]^{1/2}, 
  &\quad \tilde{Z} &= \tilde{h}(R, \tau),
  \label{eqn:kinetic_1}
\end{alignat}
\end{subequations}
where $\tilde{\Theta}_I = \tilde{\Theta}_s(R, \tilde{h}(R,\tau),\tau) 
= \tilde{\Theta}_l(R, \tilde{h}(R,\tau),\tau)$. The symmetry about $\tilde{Z}=0$
implies that $\Pd{\tilde{\Theta}_l}{\tilde{Z}} = 0$ at $\tilde{Z} = 0$.
The relevant matching conditions for the temperature in
the solid as $\tilde{Z} \to \infty$ will be discussed below.

\

The solution to this problem is now expanded as
\begin{subequations}
\begin{align}
  \tilde{\Theta}_j &= \tilde{\Theta}_j^{(0)} + \epsilon \tilde{\Theta}_j^{(1)} + O(\epsilon^2), \\
  \tilde{h} &= \tilde{h}^{(0)} + \epsilon \tilde{h}^{(1)} + O(\epsilon^2).
\end{align}
\end{subequations}
Assuming that $\epsilon^2 \hat{q} = O(\epsilon^2)$, the $O(1)$ 
solution for the temperature is straightforward to obtain and is given by
\begin{align}
  \tilde{\Theta}_l^{(0)}(R, \tilde{Z}, \tau) = 
  \tilde{\Theta}_s^{(0)}(R, \tilde{Z}, \tau)
  \equiv \tilde{\Theta}_I^{(0)}(R, \tau).
  \label{eqn:t_match}
\end{align}
The matching condition for this problem is given by
$\tilde{\Theta}_s^{(0)}(R,\tilde{Z},\tau) = \theta_s(R,0,\tau)$
as $\tilde{Z} \to \infty$. From \eqref{eqn:t_match},
we can deduce
that $\tilde{\Theta}_I^{(0)}(R, \tau) = \theta_s(R, 0, \tau)$. 
Therefore, the $O(1)$ part of the kinetic equation 
\eqref{eqn:kinetic_1} becomes
\begin{align}
  \PD{\tilde{h}^{(0)}}{\tau} = \theta_s(R, 0, \tau)
  \left[1 + \left(\PD{\tilde{h}^{(0)}}{R}\right)^2\right]^{1/2}.
  \label{eqn:kinetic_2}
\end{align}

\

Proceeding to the $O(\epsilon)$ problem, we find that the
temperatures are determined from bulk equations
\begin{align}
  \PDD{\tilde{\Theta}_j^{(1)}}{\tilde{Z}} = 0,
\end{align}
and must satisfy the Stefan condition
\begin{align}
  b^{-1} \PD{\tilde{h}^{(0)}}{\tau} = -\hat{k}\PD{\tilde{\Theta}^{(1)}_l}{\tilde{Z}}
  + \PD{\tilde{\Theta}^{(1)}_s}{\tilde{Z}}.
  \label{eqn:stefan_2}
\end{align}
By exploiting the symmetry of the problem about the
$Z$ axis, we find that the temperature in the liquid,
$\tilde{\Theta}^{(1)}_l$, must be constant in space. 
Asymptotically matching the derivatives of the solid
temperature in regions (i) and (iii) gives the relation
\begin{align}
  \PD{\theta^{(1)}_s}{\tilde{Z}} = \PD{\theta_s}{z}
  \label{eqn:deriv_match}
\end{align}
as $\tilde{Z} \to \infty$ and $z \to 0$. Using 
\eqref{eqn:deriv_match} in the Stefan condition
\eqref{eqn:stefan_2} yields
\begin{align}
  \PD{\tilde{h}^{(0)}}{\tau} = b\PD{\theta_s}{z}, \quad z = 0.
  \label{eqn:stefan_3}
\end{align}
We emphasise here that \eqref{eqn:kinetic_2} and
\eqref{eqn:stefan_3} can be treated as boundary conditions
for the problem in region (iii) away from the melt.


\subsubsection{Analysis near the melt and near the rim}

The next step is to consider the local dynamics near the
rim in order to derive an equation describing its motion. 
We switch to a travelling-wave coordinate given by
$\check{R} = (R - S(\tau)) / \epsilon^2$ and let
$Z = \epsilon^2 \check{Z}$. These scales have been chosen
in order to balance both sides of the initial interface
profile given in \eqref{eqn:ic_h}.
The position of the solid-liquid interface is written
as $h(R, \tau) =  \epsilon^2 \check{h}(\check{R})$ and
the temperatures are denoted by $\check{\Theta}_j$
for $j = l,s$. 
Upon using this scaling in \eqref{eq:nondim},
the leading-order problem in $\epsilon$ is given by
\begin{align}
  \PDD{\check{\Theta}_j}{\check{R}} + 
  \PDD{\check{\Theta}_j}{\check{Z}} 
  = 0, \quad (\check{R}, \check{Z}) \in \check{
    \Omega}_j(\tau), \quad j = s,l.
  \label{eqn:nn_bulk}
\end{align}
The Stefan condition reduces to the continuity of thermal flux 
across the interface:
\begin{align}
  \PD{\check{\Theta}_s}{\check{Z}} - \PD{\check{\Theta_s}}{\check{R}}
  \PD{\check{h}}{\check{R}} = 
  \hat{k}
  \left(
    \PD{\check{\Theta}_l}{\check{Z}} - \PD{\check{\Theta_l}}{\check{R}}
    \PD{\check{h}}{\check{R}}
  \right), \quad \check{Z} = \check{h}(\check{R}), \quad \check{R} < 0.
  \label{eqn:nn_stef}
\end{align}
The leading-order kinetic equation reads
\begin{align}
  -\OD{S}{\tau}\OD{\check{h}}{\check{R}} = 
  \check{\Theta}_I 
  \left|\PD{\check{h}}{\check{R}}\right|, \quad \check{Z}
  = \check{h}(\check{R}), \quad \check{R} < 0,
  \label{eqn:nn_kin}
\end{align}
where $\check{\Theta}_I = \check{\Theta}_s(\check{R}, \check{h}(\tau),\tau)
= \check{\Theta}_l(\check{R},\check{h}(\tau), \tau)$.
Since the thickness of the melt needs to decrease to zero as the rim
is approached, we expect that $\Pd{\check{h}}{\check{R}} < 0$ for
all $\check{R} < 0$;
therefore, the kinetic condition \eqref{eqn:nn_kin} reduces to
\begin{align}
  \PD{S}{\tau} = \check{\Theta}_I(\check{R}, \tau), \quad
  \check{R} < 0.
\end{align}
Furthermore, we have the symmetry conditions
\begin{subequations}
\begin{alignat}{3}
  \PD{\check{\Theta}_l}{\check{Z}} &= 0, 
  &\quad \check{Z} &= 0, &\quad \check{R} &< 0, \\
  \PD{\check{\Theta}_s}{\check{Z}} &= 0, 
  &\quad \check{Z} &= 0, &\quad \check{R} &> 0.
\end{alignat}
\end{subequations}
By matching to the solutions in region (ii), we obtain the 
following far-field conditions:
\begin{align}
  \check{\Theta}_l = \check{\Theta}_s \sim \ \theta_s(S(\tau),0,\tau).
\end{align}
It is straightforward to see that the bulk equations \eqref{eqn:nn_bulk}
and the Stefan condition \eqref{eqn:nn_stef}
are satisfied by temperatures that are constant in space.
Therefore, we have that
\begin{align}
  \check{\Theta}_l = \check{\Theta}_s \equiv \ \theta_s(S(\tau),0,\tau)
\end{align}
to leading order, which implies the rim moves according to
\begin{align}
  \PD{S}{\tau} = \theta_s(S(\tau),\tau).
\end{align}
The next-order problem can be used to determine the profile of the melt
near the rim; however, this is not required in the subsequent analysis. 
Finally, we note that by matching the melt heights in regions (ii) and
(iii), \emph{i.e.}, $\epsilon \tilde{h}$ and $\epsilon^2 \check{h}$,
we find that
\begin{align}
  \tilde{h}^{(0)} \sim 0, \quad R \sim S(\tau).
\end{align}
We now have all of the ingredients to write down a self-contained
problem in region (iii).


\subsubsection{A reduced model for $O(1)$ times}
In region (iii), the melt appears
to have zero thickness; it has been collapsed onto a circle
lying within the $Z = 0$ plane. The asymptotic matching into
the inner regions (i) and (ii) provides boundary conditions
on this circle. Although the melt is effectively treated as
having zero thickness, the model still captures its
evolving shape. 

\

In region (iii), the temperature field satisfies the equation
\begin{subequations}
  \label{eqn:O1_model}
\begin{align}
  \PD{\theta_s}{\tau} = \nabla^2 \theta_s + 1, \quad Z > 0,
\end{align}
with $\theta_s = 0$ when $\tau = 0$. In the far-field, we
require that $\theta_s \sim \tau$ as $|\bX| \to \infty$.
The $Z = 0$ plane is divided into two regions corresponding
to being inside and outside of the melt, $R < S(\tau)$ and
$R > S(\tau)$, respectively. For points inside of the melt,
we have Stefan and anisotropic kinetic conditions given by
(where we drop the $(0)$ subscript on $\tilde{h}^{(0)}$)
\begin{alignat}{2} \label{reduced:Stefan1}
  \PD{\tilde{h}}{\tau} &= b \PD{\theta_s}{Z}, &\quad 
  Z &= 0,\ R < S(\tau), \\
  \PD{\tilde{h}}{\tau} &= \theta_s\Bigg[1 + \bigg(\PD{\tilde{h}}{R}
  \bigg)^2\Bigg]^{1/2}, &\quad Z &= 0,\ R < S(\tau) \label{reduced:Stefan2}.
\end{alignat}
It should be noted that, because of
(\ref{reduced:Stefan1}),
the Stefan condition plays a significant r\^{o}le in this regime.
This means that the isothermal approximation fails to hold
and the melt region should no longer be expected to take a kinetic Wulff shape.

Outside of the melt, we impose a symmetry condition given by
\begin{align}
  \PD{\theta_s}{Z} = 0, \quad Z = 0,\ R > S(\tau).
\end{align}
The kinetic condition at the rim reads
\begin{align}
  \PD{S}{\tau} = \theta_s, \quad Z = 0,\ R = S(\tau).
\end{align}
Finally, it is required that
\begin{align}
  \tilde{h}(S(\tau), 0, \tau) = 0.
\end{align}
\end{subequations}
To determine asymptotically consistent
initial conditions for the position of the rim
and the profile of the solid-liquid interface, we examine the
early behaviour of \eqref{eqn:O1_model} and match it to the
small-time solution given by \eqref{eq:a:appelipse}.


\subsubsection{Early behaviour of model for $O(1)$ times}
\label{subsec:stcod2}

The relevant scaling to resolve the early time behaviour and match
into the small-time regime is given by
$\tau = \epsilon^{1/2} \hat{\tau}$, $\theta_s = \epsilon^{1/2} \hat{\theta}_s$,
$\bX = \epsilon \hat{\bX}$, $\tilde{h} = \epsilon\hat{h}$, and
$S = \epsilon \hat{S}$. From the leading-order problem in $\epsilon$, it is
straightforward to deduce that $\hat{\theta} = \hat{\tau}$.
The leading-order kinetic conditions that hold within the melt
and at the rim are then given by
\begin{subequations}
\begin{align}
  \PD{\hat{h}}{\hat{\tau}} &= \hat{\tau}\left[1 + \bigg(\PD{\hat{h}}{
      \hat{R}}\bigg)\right]^{1/2}, 
  \label{eqn:st_h}\\
  \PD{\hat{S}}{\hat{\tau}} &= \hat{\tau}.
  \label{eqn:st_S}
\end{align}
\end{subequations}
From \eqref{eq:a:appelipse}, we see that in the small-time regime,
the rim grows like $\tau^2 / 2 + O(\epsilon)$ for $\tau \sim O(1)$;
therefore, we can solve \eqref{eqn:st_S} and by matching we obtain
$\hat{S}(\hat{\tau}) = \hat{\tau}^2 / 2$. The solution for $\hat{S}$
motivates seeking a similarity solution to \eqref{eqn:st_h} of
the form $\hat{h} = \hat{\tau}^2 \hat{H}(\hat{R} / \hat{\tau}^2)$. 
Using this ansatz in \eqref{eqn:st_h} gives the problem
\begin{align}
  2 \left[ \hat{H}(\zeta) - \zeta \hat{H}'(\zeta)\right]
  = \left[1 + (\hat{H}'(\zeta))^2\right]^{1/2},
\end{align}
where $\zeta = \hat{R}/\hat{\tau}^2$ and $\hat{H}$ satisfies
$\hat{H}(1/2) = 0$. The solution is
$\hat{H}(\zeta) = A (1/4 - \zeta^2)^{1/2}$ or, equivalently,
\begin{align}
  \hat{h} = A \left(\frac{\hat{\tau}^4}{4}- \hat{R}^2\right)^{1/2},
\end{align}
where $A = 1$ is a constant that can be determined by matching
to \eqref{eq:a:appelipse} as $\tau \sim O(1)$.
From this analysis, we can conclude that the model in \eqref{eqn:O1_model}
should have initial conditions for the interface given by
\begin{align}
  \tilde{h} \sim \left(\frac{\tau^4}{4} - R^2\right)^{1/2},
  \quad
  S \sim \frac{\tau^2}{2}
  \label{eqn:sim}
\end{align}
as $\tau \sim 0$. For $0 < \tau \ll 1$, \eqref{eqn:sim} describes the
early growth of the melt in the $O(1)$ time regime, which is consistent
with the long-term growth in the first time regime. 

\


\subsection{Linear Stability for Times of O(1)}  \label{subsec:stability}

We now examine the linear stability of the system 
using the reduced model \eqref{eqn:O1_model}.
The calculation involves two main steps. First, a
base state corresponding to a growing
axisymmetric melt is computed. Finally, we determine
the growth rates of small, azimuthally varying
perturbations to the base state. Our analysis will
focus on constructing local solutions valid near, 
but not too close to, the rim. 

\

Our calculation of the base state begins by
introducing a travelling wave coordinate $\breve{X}$ such
that $\breve{X} = R - S(\tau)$ and letting $\breve{Z} = Z$.
We focus on the local behaviour of solutions near the 
rim so that $\breve{X}^2 + \breve{Z}^2 \ll 1$. The temperature
and the melt thickness are written as 
$\theta_s \sim \breve{\theta}_s(\breve{X}, \breve{Z})$ and $h \sim \breve{h}(\breve{X})$,
where we expect from \eqref{eqn:ic_h} that
$\breve{h}(\breve{X}) \sim \breve{h}_1(-\breve{X})^{1/2}$ for
sufficiently small $\breve{X}$. 

\

Close to the rim, the temperature $\breve{\theta}_s$ 
approximately satisfies Laplace's 
equation:
\begin{align}
  \PDD{\breve{\theta}_s}{\breve{X}} + \PDD{\breve{\theta}_s}{\breve{Z}} = 0.
\end{align}
The Stefan and kinetic conditions read
\begin{alignat}{2}
  -\PD{S}{\tau}
  \PD{\breve{h}}{\breve{X}} &= 
  b \PD{\breve{\theta}_s}{\breve{Z}}, &\quad \breve{Z} &= 0,\ \breve{X} < 0, 
  \label{eqn:ls_stef}
  \\
  -\PD{S}{\tau}
  \PD{\breve{h}}{\breve{X}} &= 
  \breve{\theta}_s\Bigg[1 + \left(\PD{\breve{h}}{\breve{X}}\right)^2\Bigg]^{1/2},
  &\quad \breve{Z} &= 0,\ \breve{X} < 0,
  \label{eqn:ls_kin}
\end{alignat}
respectively. The symmetry condition is given by
\begin{align}
  \PD{\breve{\theta}_s}{\breve{Z}} = 0, \quad \breve{Z} = 0,\ \breve{X} > 0.
  \label{eqn:ls_sym}
\end{align}
and the rim evolves according to
\begin{align}
  \PD{S}{\tau} = \breve{\theta}_s, \quad \breve{Z} = 0,\ \breve{X} = 0.
  \label{eqn:ls_st}
\end{align}
Since we have assumed that $\breve{\theta}_s$ is independent of $\tau$,
we immediately deduce from \eqref{eqn:ls_st} that the rim moves with a 
constant velocity, $V$, given by $V = \breve{\theta}_s(0,0)$.

\

\begin{figure}
  \centering
  \includegraphics[width=0.5\textwidth]{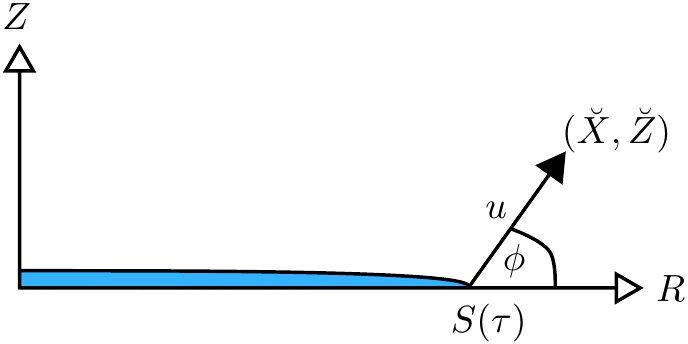}
  \caption{A schematic diagram of the local polar coordinates 
    given by $\breve{X} = u \cos \phi$ and $\breve{Z} = u \sin \phi$
    that are centred at the rim. It is convenient to write the local
    temperature profile in terms of these coordinates; see text for details.}
  \label{fig:local}
\end{figure}

An approximate solution for the temperature can be obtained by
converting to local polar coordinates that are centred at the
rim. Thus, we introduce the change of variable
\begin{align}
  \breve{X} = u \cos \phi, \quad \breve{Z} = u \sin \phi,
  \label{eqn:ls_polar}
\end{align}
where $u$ is the local radius and $\phi$ is the polar angle
measured relative to the positive $\breve{X}$ axis; see 
Fig.~\ref{fig:local} (a). 

An approximate solution
for the temperature can be written as
\begin{align}
  \breve{\theta}_s \sim V + \breve{\theta}_1 u^{1/2} \cos(\phi / 2) + \gamma \breve{X},
  \label{eqn:ls_theta_b}
\end{align}
which satisfies the symmetry condition \eqref{eqn:ls_sym} and
where $\breve{\theta}_1$ is a constant that can be determined from
the Stefan condition \eqref{eqn:ls_stef}. In particular, by inserting
\eqref{eqn:ls_theta_b} in \eqref{eqn:ls_stef} and using the fact that
$\breve{h} \sim \breve{h}_1 (-\breve{X})^{1/2}$ for $\breve{X} \sim 0^{-}$, we find
\begin{align}
  \frac{1}{2}\,V \breve{h}_1 (-\breve{X})^{-1/2} \sim \frac{1}{2}\,\breve{\theta}_1(-\breve{X})^{-1/2},
  \quad \breve{X} \sim 0^{-}
\end{align}
so that $\breve{\theta}_1 = V \breve{h}_1$. 
Using a similar procedure in the kinetic
condition \eqref{eqn:ls_kin} shows that $\breve{h}_1 = 1$. 
The parameter $\gamma$
is taken to be a free parameter and we will investigate the role it plays
in controlling the stability of the problem.

We now investigate the stability of the base state by adding 
small perturbations of order $\delta \ll 1$ to $\breve{\theta}_s$
and $S$. To simplify matters, we suppose that we are looking
locally near $(X, Y, Z) = (V \tau + \breve{X}, \breve{Y}, \breve{Z})$,
where $\breve{X}^2 + \breve{Y}^2 + \breve{Z}^2 \ll 1$,
and can consider the rim as a straight line on these scales. Taking the
rim to be flat is reasonable when the perturbation wavenumber in the azimuthal
direction is large. Note that $\breve{X} = \breve{Y} = \breve{Z} = 0$
corresponds to a point on the base-state rim and, thus, we have effectively
attached a Cartesian coordinate system to this point. 
We write the local temperature and the position of the rim as
\begin{subequations}
\label{eqn:ls_pert}
\begin{align}
  \theta_s &\sim \breve{\theta}_s(\breve{X},\breve{Z})
  + \delta \breve{\Theta}_s(\breve{X},\breve{Z})\,\ee^{\ii \kappa
    \breve{Y} + m \tau}, \\
  S &\sim V \tau + \delta \breve{S}\,\ee^{\ii \kappa \breve{Y} + m\tau},
\end{align}
\end{subequations}
where  $\kappa$ and $m$ denote the wavenumber and  growth rate of the
perturbations, respectively, 
and $\breve{\theta}_s$ is given by \eqref{eqn:ls_theta_b}.
The perturbation to the temperature satisfies the equation
\begin{align}
  \PDD{\breve{\Theta}_s}{\breve{X}} + \PDD{\breve{\Theta}_s}{\breve{Z}}
  -\kappa^2 \breve{\Theta}_s = 0,
\end{align}
together with
\begin{align}
  \PD{\breve{\Theta}_s}{\breve{Z}} = 0, \quad \breve{Z} = 0,\ \breve{X} > 0.
\end{align}
The solution can be found using the local polar coordinates in
\eqref{eqn:ls_polar} and written as
\begin{align}
  \breve{\Theta}_s = \breve{\Theta}_1 u^{-1/2} \ee^{-\kappa u}\cos(\phi/2),
\end{align}
where $\breve{\Theta}_1$ is a constant that is to be determined. 
An equation governing the perturbation to the rim position can be derived
from inserting \eqref{eqn:ls_pert} into the kinetic condition
$\Pd{S}\tau = \theta_s(S(\tau),\tau)$, expanding about $\delta \ll 1$, and
taking the $O(\delta)$ part:
\begin{align}
  m \breve{S} = \PD{\breve{\theta}_s}{\breve{X}}\,\breve{S}
  + \breve{\Theta}_s, 
  \quad \breve{Z} = 0,\ \breve{X} = 0.
  \label{eqn:ls_kin_p}
\end{align}
We note that
\begin{align}
  \PD{\breve{\theta}_s}{\breve{X}} \sim
  \frac{V}{2}\,\breve{X}^{-1/2} + \gamma, \quad
  \breve{\Theta}_s \sim \breve{\Theta}_1 \breve{X}^{-1/2}
\end{align}
as $\breve{X} \sim 0$ and $\breve{Z} = 0$, both of which become singular 
as $\breve{X} \to 0$. In order for
the kinetic condition \eqref{eqn:ls_kin_p} to remain well defined, we need
$\breve{\Theta}_1 = -(V/2)\breve{S}$, which yields
\begin{align}
  m \breve{S} = \gamma \breve{S},
\end{align}
\emph{i.e.}, the perturbation growth rate $m$ is exactly equal to
the parameter $\gamma$ in the base-state temperature profile
\eqref{eqn:ls_theta_b}. 
This linear analysis thus indicates instability if
there is a background temperature gradient in the direction
of propagation of the rim, $\gamma>0$,
but stability for a negative gradient, $\gamma<0$.
Note that in the case of instability, the
growth rate of the perturbations is independent of
the wave number, in contrast
to unstable Hele-Shaw or Stefan problems without surface
tension/energy, where growth rate increases with
wave number and can be arbitrarily high.
Note that similar stability results for another
free boundary problem were obtained in
Howison {\it et al}.~\cite{HOW}.

\

Given the absence of exact and of approximate long-time
solutions about which to perturb,
it is not immediately apparent what values
$\gamma$ might take in practice. Intuitively we might expect
$\gamma$ to be positive,
since melting at the interface has the effect of locally reducing temperature,
at least for relatively low times $\tau$. The simulations by
Hennessy \cite{hennessy2010} support this claim, although they do
not consider heat transfer in the axial direction.
If $\gamma$ is positive, we then expect a mild instability
whose form will also be influenced by any further anisotropy,
for instance, the usual six-fold one in the $(X,Y)$ plane.


\subsection{Other Anisotropy Functions}  \label{subsec:otheranisot}

We now briefly outline the results that are obtained for the
anisotropy functions \eqref{eqn:f2} and \eqref{eqn:f3}. 
Full details about the solutions in the early-time regime
and the solution of Charpit's equations are given in
Appendix \ref{app:charpit}. 

\subsubsection{Dynamics with $f(\psi) = \epsilon + \sin^2 \psi$}
\begin{figure}
  \centering
  \includegraphics[width=0.8\textwidth]{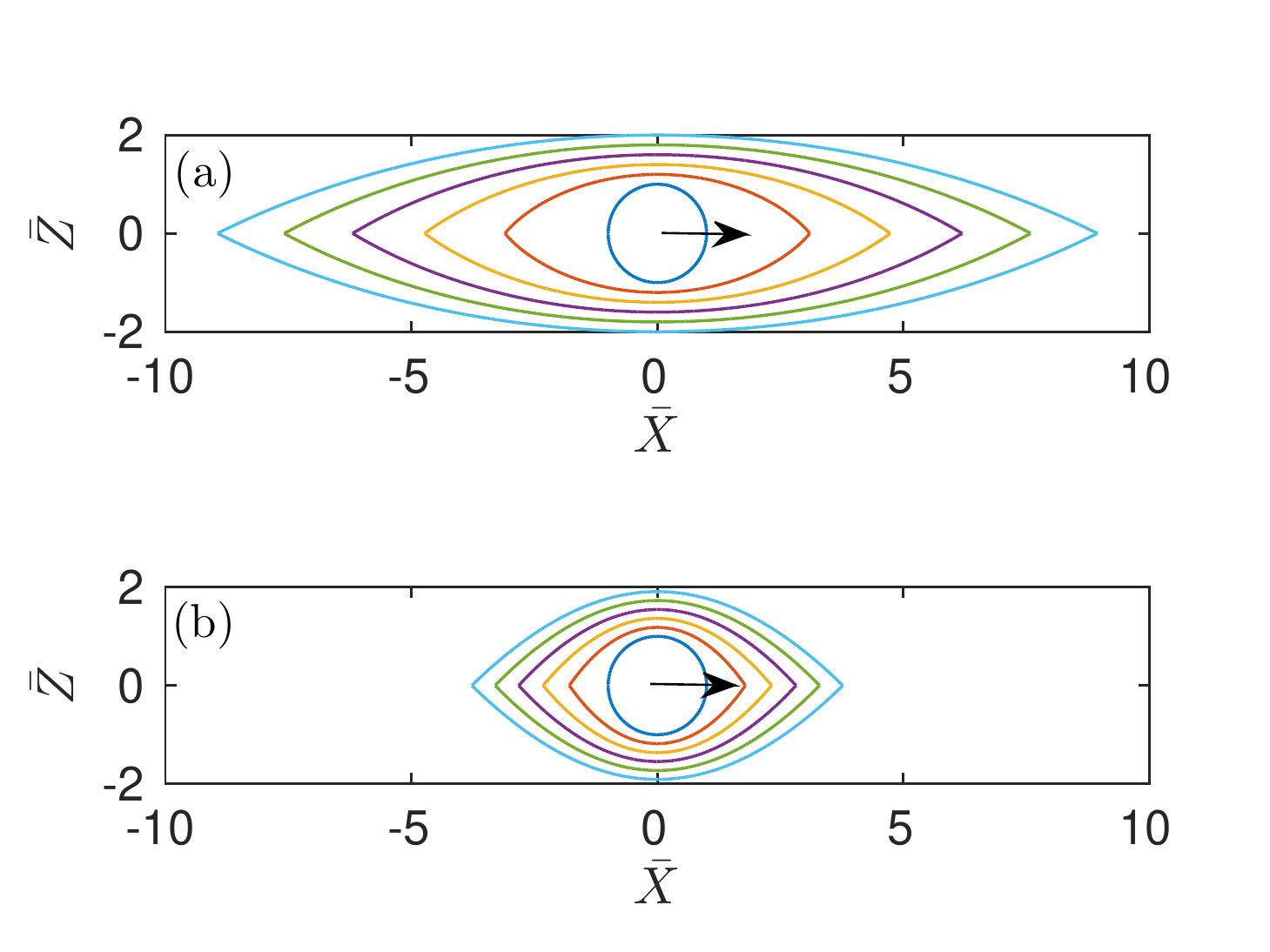}
  \caption{Early-time evolutions of an initially spherical
    melt for anisotropy functions
    $f(\psi) = \epsilon + \sin^2 \psi$ (panel a) and 
    $f(\psi) = \epsilon / (1 + \epsilon - \sin^2 \psi)$ (panel b) when 
    $\epsilon = 0.1$. The curves in panels (a) and (b) are obtained
    from the solutions \eqref{eqn:char_2} and \eqref{eqn:char_3}, 
    respectively. 
    The positions of the
    interface are shown at equally spaced values of $\bar{s}$ given by
    $\bar{s} = 0$, $2$, $4$, $6$, $8$, and $10$, corresponding to
    rescaled dimensionless times $\bar{\tau} = 0$, $2$, $2.83$, $3.46$,
    $4.0$ and $4.47$, respectively.  Both anisotropy functions lead to the
    formation of a corner, and this happens when $\bar{\tau} = 1.49$ 
    in panel (a) and $\bar{\tau} = 0.32$ in panel (b).  
    As $\bar{\tau}\to\infty$,
    the interface profiles approach the kinetic Wulff shapes shown in 
    Fig.~\ref{fig:kin_wulff}.}
  \label{fig:anisotropy_2_3}
\end{figure}

In the early-time regime given by $\tau = O(\alpha^{1/2})$, the solid-liquid
interface can be written parametrically as
\begin{align}
X = [\alpha + s(1 + \epsilon + \sin^2 \varphi )] \cos\varphi,
\quad
Z = [\alpha + s(\epsilon - \cos^2 \varphi )]\sin\varphi,
\label{eqn:char_2}
\end{align}
where $\varphi \in [0, 2\pi )$ and $s = \tau^2 / 2$. 
Interface profiles at various times are shown in 
Fig.~\ref{fig:anisotropy_2_3} (a). The interface remains smooth
until $s = \tau^2 / 2 = \alpha / (1 - \epsilon)$, at which point
a corner develops at the rim due to intersecting characteristics.
The early growth of the rim
for $s / \alpha \ll \epsilon$ scales like $S \sim \tau$; however,
the longer-term growth of the rim for $s / \alpha \gg \epsilon$ is
reduced by the corner and we find that 
$S \sim \epsilon^{1/2} \tau^2$. The thickness of the melt in the
axial direction grows like $\epsilon \tau^2$ for all times. 

\

For larger times, the separation of length scales in the radial
and axial directions can, in principle, be exploited and 
the model can be reduced using a similar analysis to that
in Sec.~\ref{subsec:orderone}. However, the current model is
expected to require
additional mechanisms such as surface energy to 
act to regularise the corner. 
Therefore, we do not proceed with the model reduction in this
case. Nevertheless, we note that because of the slower radial
growth for early times, in getting to terms to balance in
a model equivalent to (\ref{eqn:O1_model}), larger time and temperature
scalings are needed: $\tau = \epsilon^{-1/4}\tau^*$ and
$\theta = \epsilon^{-1/4}\theta^*$. 


\subsubsection{Dynamics with $f(\psi) = \epsilon / (1 + \epsilon
  - \sin^2 \psi)$}

In this case, the position of the interface in the early-time
regime, $\tau = O(\alpha^{1/2})$, is given by
\begin{align}
  X = \left[ \alpha + \frac {\epsilon s(3\sin^2 \varphi + \epsilon)}
    {(\sin^2\varphi + \epsilon)^2} \right] \cos\varphi,
  \quad 
  Z = \left[ \alpha - \frac {\epsilon s(3\cos^2 \varphi -1-\epsilon)}
{(\sin^2\varphi + \epsilon)^2} \right] \sin\varphi,  
\label{eqn:char_3}
\end{align}
where, again, $s = \tau^2 / 2$. Fig.~\ref{fig:anisotropy_2_3} (b)
shows the corresponding interface profiles at various times. 
Here, the corner appears in the very
early stages of melt growth, in particular, when 
$s = \tau^2 /2 = \epsilon a / (2 - \epsilon)$. The 
growth of the rim scales like 
$S \sim \alpha + (\alpha \epsilon)^{1/2}\tau$ for $s \ll \alpha / \epsilon$
and like $S \sim \epsilon \tau^2$ for $s \gg \alpha / \epsilon$. 
The axial growth scales like $\epsilon \tau^2$ for all time. 

\

For this particular anisotropy, the eventual growth of the melt
both parallel and normal to the $c$ axis is the same order of
magnitude. The aspect ratio of the melt roughly approaches
5:2 and, therefore, it is not possible to
simplify the model for $O(1)$ times. 

\

\subsubsection{Commonalities of the early-time growth}

The three anisotropy functions that we consider produce
interface profiles with common growth features in the
early-time regime. For instance, all three cases lead to
melts that evolve into their kinetic Wulff shapes given
by \eqref{eqn:kin_wulff}. In fact, an analysis for arbitrary
anisotropy functions in Appendix \ref{app:charpit} shows this
will always be the case. Furthermore, the growth of the melt
in the axial direction, \emph{i.e.} along the $c$ axis,
is always found to be quadratic with time. As shown in
Appendix \ref{app:nucleation}, if nucleation occurs
much later than when the system is irradiated, then
the axial growth becomes linear for all anisotropy functions.


\section{Discussion and Conclusion}
\label{sec:Discuss}


In this paper, we have formulated and analysed a mathematical
model describing the anisotropic growth of a Tyndall figure
into a crystal of superheated ice. Both the solid and liquid phases are
assumed to be volumetrically heated by the absorption of incoming
radiation, which drives the melting
process. The anisotropic growth of the Tyndall figure
is a result of the molecularly smoothly basal planes of the ice crystal
melting at a much slower rate in comparison to molecularly 
rough prism planes. This phenomenon 
is modelled using a kinetic coefficient
that depends on the orientation of the solid-liquid interface. 
The relationship between the kinetic coefficient
and the crystal orientation is quantified through an
anisotropy function. 
Our analysis indicates that there are two key time regimes
for the melt evolution.  The first of these describes the
rapid initial growth of the Tyndall figure into its kinetic
Wulff shape due to volumetric heating. The second time
regime describes the slower, diffusion-dominated growth.

\

The problem in the first time regime amounted to solving an
anisotropic Eikonal equation. Remarkably, it was possible
to obtain an analytical solution to this equation for an arbitrary 
anisotropy functions. Using this solution, we examined
the interface profiles and kinetic Wulff shapes that are 
obtained for three different anisotropy functions.
These anisotropy functions led to
a rich variety of melt shapes including long rectangles
with rounded ends, oblate spheroids, as well as
thick and thin lenses. Qualitatively, we found that the
smoothest melts and the smallest aspect ratios
occur when the anisotropy function has broad
maxima; anisotropy functions with narrow maxima gave rise to corners
and lens-shaped melts that can have order-one aspect ratios.
Regardless of the anisotropy function, the
thickness of the melt in the direction of the $c$ axis
was found to grow quadratically with time. This is in 
contrast to the radial growth parallel to the basal planes,
which was highly dependent on the anisotropy function. 
These findings have important practical implications,
as they suggest that experimental data for the radial growth of
the melt can aid in the determination of likely candidates for the
anisotropy function. This is not the case for axial growth,
which is predicted to be roughly the same for all anisotropy functions.

\

By exploiting the thin aspect ratio of the melt figure, we showed that
a simplified model for the evolution in the second time regime can 
be derived by systematically collapsing the
three-dimensional melt figure to a two-dimensional surface
with zero thickness along the axial direction. 
This model was then used to carry out a linear stability analysis,
the results of which suggest that an instability
will occur if the temperature field locally increases in the
direction of radial growth. Such an instability would likely
lead to fingers and could drive the formation of a Tyndall star
similar to that shown in Fig.~\ref{fig:exp}.

\

The results from our analysis, in combination with the experimental
observations by Takeya \cite{T}, may give some insight into appropriate
anisotropy functions for the melting of ice crystals. In particular, the melts
documented by Takeya have a remarkably constant thickness in direction
of the $c$ axis which diminishes relatively rapidly near the rim. In 
addition, the aspect ratio of the melt is small and on the order of
1:10. These observations suggests that an appropriate anisotropy function
for modelling the growth of Tyndall figures would be similar to
that in \eqref{eqn:f} but with much broader maxima at
$\psi = \pm \pi/2$.

\

Further predictions about Tyndall star evolution can be accessed
through numerical simulations of our model. From a computational
perspective, simplified models such as \eqref{eqn:O1_model} 
are advantageous due to the reduced dimensionality of the free
boundary and are relatively straightforward to implement. 
Numerical simulations of such a model can provide insights into
when the condition for instability is satisfied and offer a means
of probing nonlinear melt morphologies. Furthermore, such simulations
could explore whether the onset of instability is linked
to growth along the $c$ axis, which has been suggested by
experimental studies \cite{M,T}.
Thus, there is a wide range of exciting and unanswered problems relating to
the formation and evolution of Tyndall stars, and we hope this work 
not only provides some of the foundations that can aid in tackling these, 
but also motivation for doing so.


\section*{Acknowledgements}

The authors wish extend their sincerest gratitude to John Ockendon for bringing the interesting problem of Tyndall stars to their attention, countless enthusiastic discussions on the subject, and his very generous guidance during the development of this manuscript. 



\appendix
\section{Solution of Charpit's Equations for the Anisotropic Eikonal Equations}     \label{app:charpit}


An asymptotic analysis of the model revealed that the early-time
interface profiles can be obtained by
solving an anisotropic Eikonal equation of the form
\begin{subequations}
  \label{app:eikonal}
\begin{align}
  (s_X^2 + s_Z^2)^{1/2}\hat{f}\left(s_X(s_X^2 + s_Z^2)^{-1/2}\right) = 1,
  \label{app:eikonal_pde}
\end{align}
where $s_X = \partial s / \partial X$ and $s_Z = \partial s / \partial Z$.
Equation \eqref{app:eikonal} is supplemented with the
condition 
\begin{align}
  s_0 = s(X_0, Z_0) \equiv 0, \quad  X_0^2 + Z_0^2 = \alpha^2.
  \label{app:eikonal_ic}
\end{align}
\end{subequations}
The solution to this problem can be obtained using Charpit's equations,
which generalise the well-known method of characteristics to fully
nonlinear first-order hyperbolic partial differential equations (PDEs)
\cite{OHLM}. 
We recall that when applying the method of characteristics, one must 
simultaneously solve for the characteristic directions and the solution to
the PDE on these characteristics. 
The idea behind Charpit's method
is to treat the first derivatives of the solution to the PDE
as additional unknowns that must be found along the characteristic directions. 
Thus, when applying Charpit's method to this problem,
we must simultaneously solve for
the characteristic directions, $X$ and $Z$, as well as the solution
$s$ and its derivatives $s_X$ and $s_Z$ along the characteristics. Although
these five unknowns are effectively treated as independent variables, Charpit's
equations ensure that they always vary in a consistent manner. 

To apply Charpit's method to \eqref{app:eikonal}, we first let
$p = s_X$, $q = s_Z$, and we write the PDE in \eqref{app:eikonal_pde} as
\begin{align}
  G(X,Z,s,p,q) = (p^2 + q^2)^{1/2} \hat{f}\left(p(p^2 + q^2)^{-1/2}\right) - 1
  \equiv 0.
  \label{app:G}
\end{align}
The condition in \eqref{app:eikonal_ic} can be
treated as initial data and parametrised according to
\begin{subequations}
\label{app:char_ic}
\begin{alignat}{2}
  &s_0(\varphi) = s(X_0(\varphi), Z_0(\varphi)) = 0, &\quad \zeta &= 0, \\
  &X_0(\varphi) = \alpha \cos \varphi, &\quad \zeta &= 0, \\
  &Z_0(\varphi) = \alpha \sin \varphi, &\quad \zeta &= 0,
\end{alignat}
where $\varphi \in [0, 2\pi)$
and $\zeta$ is an arbitrary parameter that measures distance along each
characteristic direction. 
Initial conditions for $p$ and $q$, given by
$p_0$ and $q_0$, can be obtained by (i) differentiating the condition
$s_0(\varphi) = s(X_0(\varphi), Z_0(\varphi)) \equiv 0$ with respect
to $\varphi$ and (ii) requiring the PDE \eqref{app:G} to hold on
the initial curve, $G(X_0, Z_0, s_0, p_0, q_0) \equiv 0$. By
simultaneously solving two these equations, we obtain
\begin{align}
  p_0(\varphi) = \frac{\cos \varphi}{\hat{f}(\cos \varphi)}, \quad
  q_0(\varphi) = \frac{\sin \varphi}{\hat{f}(\cos \varphi)}, \quad
  \zeta = 0.
\end{align}
\end{subequations}
Charpit's equations for this problem can be written as
\begin{subequations}
\begin{align}
  \dot{X} &= \PD{G}{p}, \\
  \dot{Z} &= \PD{G}{q}, \\
  \dot{s} &= 1, \\
  \dot{p} &= 0, \\
  \dot{q} &= 0,
\end{align}
\end{subequations}
where the dot denotes differentiation with respect to $\zeta$.
Upon solving these equations with the initial conditions in
\eqref{app:char_ic}, we find that $s \equiv \zeta$, so that
$\zeta$ can be replaced by $s$. In addition, 
we have $p \equiv p_0$, $q \equiv q_0$, and 
\begin{subequations}
  \label{app:eq:chargen}
  \begin{align}
    X &=  [\alpha + s\fh (\cos\varphi)] \cos\varphi +
    s\fh '(\cos\varphi) \sin^2 \varphi, \\
    Z & =  [\alpha + s(\fh (\cos\varphi) - \fh '(\cos\varphi))]
    \sin\varphi,
  \end{align}
\end{subequations}
with the prime denoting derivative with respect to argument.

\

For the anisotropy function (a) $f(\psi) = (\epsilon^2 +
\sin^2\psi)^{1/2}$, we have that
$\fh (w) = (\epsilon^2 + w^2)^{1/2}$.
After inserting this expression into \eqref{app:eq:chargen}
and some algebra, the solution can be written as
\begin{align}
  X = \left( \alpha + \frac{s(1 + \epsilon^2)}{(\epsilon^2 + \cos^2 \varphi)^{1/2}}
  \right) \cos\varphi,
  \quad
  Z = \left( \alpha + \frac{s \epsilon^2}{(\epsilon^2 + \cos^2 \varphi)^{1/2}} \right)
\sin\varphi.
\label{app:eq:earlyshape:a1}
\end{align}
The properties of this solution are described in Sec.~\ref{subsec:early}.
For the anisotropy functions (b) $f(\psi) = \epsilon + \sin^2\psi$
and (c) $f(\psi) = \epsilon/(1+\epsilon-\sin^2\psi)$,
we find that
\begin{align}
X = [\alpha + s(1 + \sin^2\varphi + \epsilon)] \cos\varphi, \quad
Z = [\alpha - s(\cos^2\varphi - \epsilon)] \sin\varphi,
\label{app:eq:earlyshape:b1}
\end{align}
and 
\begin{align}
X = \left[ \alpha + \frac {\epsilon s(3\sin^2 \varphi + \epsilon)}
{(\sin^2\varphi + \epsilon)^2} \right] \cos\varphi, \quad
Z = \left[ \alpha - \frac {\epsilon s(3\cos^2 \varphi -1-\epsilon)}
{(\sin^2\varphi + \epsilon)^2} \right] \sin\varphi,
\label{app:eq:earlyshape:d1}
\end{align}
respectively. These solutions with $\epsilon = 0.1$
are shown in Figs.~\ref{fig:anisotropy_1}
and \ref{fig:anisotropy_2_3}.

\

For case (b), focusing on that part of the free boundary lying
in the first quadrant, $0 \le \varphi \le \pi / 2$,
we see that
some of the characteristics are directed down, towards $Z=0$,
and intersection of characteristics starts, on the $X$ axis,
when $s = \alpha/(1 - \epsilon) \sim a$ at
$X = \alpha + \alpha(1 + \epsilon)/(1 - \epsilon) = 2\alpha/(1 - \epsilon)
\sim 2a$. For later times this method of characteristics
indicates multiple-valued solutions.
To avoid this, the convex part of the curve is taken,
giving corners on $Z=0$ for $s > \alpha/(1 - \epsilon)$.
These would be expected to be rounded off by any sort of
surface-tension or surface-energy effects so that a
Gibbs--Thomson term is introduced into the free-boundary
conditions. A mathematically simpler way of regularising the
problem would be to replace the anisotropic Eikonal
equation, which is a first-order flow, by a
mean-curvature flow. Results of Barles \& Souganidis
\cite{BS} could
be applied to give continuous dependence of solutions
on the coefficient of any curvature term included
in \eqref{eqn:kin1}. This would again
indicate that we should get the interface by
taking the convex part of the curve.

\

The same corner formation is seen for the anisotropy function
(c). In this case, the corner forms very quickly, when
$s = \epsilon \alpha/(2-\epsilon)$, and close to the initial
free boundary, at $X = 2\alpha/(2-\epsilon)$.

\

The range of possible short-time interface
behaviour is large because
the growth in the $X$ direction can have quite different
qualitative behaviour. For case (b), with $s \ll \alpha/\epsilon$,
$X = 2s(\epsilon + \alpha/s)^{1/2} \sim 2(\alpha s)^{1/2}
= (2\alpha)^{1/2} \tau$ so that the growth is only linear in time.
The final case (c), has $\cos^2 \varphi \sim 1
- (2\epsilon s/\alpha)^{1/2}$ and $X \sim \alpha (1 +
(2\epsilon s/\alpha)^{1/2}) = \alpha (1 + (\epsilon/\alpha)^{1/2}\tau)$.

\

For large times, in the sense of $s \gg \alpha/\epsilon$, 
the behaviour of $Z$ is the same for all three anisotropy functions:
$Z \sim \epsilon s \sim  \epsilon \tau^2 / 2$. 
However, the long-time
growth in the $X$ direction is reduced, thanks to the
appearance of the corner. For (b), the corner's position
is, in general, given by $X = [\alpha + s(1 + \sin^2\varphi
+ \epsilon)] \cos\varphi$ with $Z = [\alpha - s(\cos^2\varphi
- \epsilon)] \sin\varphi = 0$. Since $0 < \varphi < \pi/2$,
this gives $\cos^2 \varphi = \epsilon + \alpha/s$ and
$X = 2s(\epsilon + \alpha/s)^{1/2} \sim 2\epsilon^{1/2} s
= \epsilon^{1/2} \tau^2$
for $s \gg \alpha /\epsilon$. 
Very similar calculations for (c) show that
the corner location can be obtained implicitly
from 
\begin{align}
\frac X\alpha \sim \frac {2\cos\varphi}{3\cos^2\varphi - 1},
\quad \frac {\epsilon s}{\alpha} \sim
\frac {(1 - \cos^2\varphi)^2}{3\cos^2\varphi - 1},
\label{eq:rimd:implicit}
\end{align}
for $0 < \varphi < \cos^{-1} (1 / \sqrt{3})$.
For $s \gg \alpha/\epsilon$, this gives $\cos^2\varphi \sim (1/3)$
and we get $X \sim 3^{3/2} / 2 \epsilon s =
(3^{3/2}/4) \epsilon\tau^2$.

\

By taking the modified time variable $s$ sufficiently large
in comparison to $\alpha$ in \eqref{app:eq:chargen},
the longer-term interface profile for an arbitrary anisotropy function
is given by
\begin{subequations}
\label{app:eq:charuni}
\begin{align}
  X / s&\sim \fh (\cos\varphi) \cos\varphi +
\fh '(\cos\varphi) \sin^2 \varphi, \\
Z / s &\sim \fh (\cos\varphi)\sin\varphi - \fh '(\cos\varphi) \cos\varphi
\sin\varphi,
\end{align}
\end{subequations}
\underline{independent}
of the initial shape. Equation \eqref{app:eq:charuni}
is, in fact, equivalent to \eqref{eqn:kin_wulff} and therefore,
the interface profiles approach the kinetic Wulff shapes. 
The direction of the
characteristics, $Z/X$, can be differentiated with respect
to $\varphi$ to check if this ever decreases, leading to
corner formation from an initially convex shape. After
some manipulation, the derivative turns out to be
\[
\left( \fh - \OD {}\varphi (\fh ' \sin\varphi) \right) \fh =
\left( f + \OD {^2 f}{\psi^2} \right) f \, .
\]
The criterion for a continued smooth interface is then
$f + \Od {^2 f}{\psi^2} \geq 0$. 
Cahoon \emph{et al}.~\cite{CMW} and 
Wettlaufer \emph{et al}.~\cite{WJE} find the same basic law
for interface motion gives the rate of
change $\Od\kappa s = (f + \Od {^2 f}{\psi^2}) \kappa^2$
for the interface curvature $\kappa$.
The same key
combination appears in curvature-flow models for
phase change with significant Gibbs--Thomson effect
\cite{AG, G}. In these works, the $(f + \Od {^2 f}{\psi^2})$ term
multiplies curvature in the velocity law and, to avoid
negative diffusion, all angles making $(f + \Od {^2 f}{\psi^2})$
positive are prohibited, leading to corners in the interface
for all positive time.


\section{The Role of Nucleation}  \label{app:nucleation}

We now give a brief discussion of the effect of surface
energy in the nucleation process, while still neglecting the
air bubble that appears in the melt. We concentrate
on the implications of the balance between the superheat
temperature and the local equilibrium temperature
for a spherical liquid body of a given size; Chadham 
\emph{et al.}~\cite{CHO} discusses related  effects
in the growth of crystals when
the Gibbs--Thomson effect is the only stabilising
action.

We suppose that nucleation occurs when the temperature in the solid
exceeds a nucleation temperature $T_n$ given 
by the Gibbs--Thomson relation
\begin{align}
  T_n = T_0\left(1 + \frac{2 \gamma}{\rho L a_n}\right),
  \label{app:gt}
\end{align}
where $\gamma$ is the interface energy,
$a_n$ is the nucleation radius.
The time at which nucleation occurs,
measured relative to the moment the system is
irradiated, is denoted by $t_n$. Before
nucleation occurs, the temperature in the 
solid increases like 
$T_s = T_0 + q_s t / (\rho c_{ps})$;
therefore, the nucleation time and temperatures
can be related via 
$t_n = \rho c_{ps} (T_n - T_0)/ q_s$.

So far we have been assuming that the nucleation
temperature is close to the bulk melting temperature,
$T_n \simeq T_0$, so that nucleation immediately
occurs upon irradiation, resulting in
an initial liquid-solid interface
that is approximately a sphere of radius
$a$, which is small compared to $\epsilon \ell$.
The condition $a \ll \epsilon \ell $ allows the
melt to become a developed spheroid when the 
dimensionless time
$\tau$ is $O(1)$ in size. Note that if $\epsilon \ll a \ll 1$ there
is a significant change to Sec.~\ref{subsec:orderone},
with the Tyndall figure no longer being of thickness
order $\epsilon$.

We now consider the opposite case whereby 
$T_n \gg T_0$ so that nucleation occurs much later than
when the system is irradiated. The bulk
temperatures in this case will be large during the 
early evolution of the melt and will influence its
growth kinetics. To study the behaviour in this 
late-nucleation regime, we non-dimensionalise
\eqref{eq:field:dim}--\eqref{eq:kinetic:dim}
by writing $\bx = \bar{\ell}\bar{\bX}$,
$t = t_n + (\bar{\ell}^2 / \kappa_s) \bar{\tau}$, and
$T = T_0 + \overline{\Delta T} \bar{\theta}$,
where $\overline{\Delta T} = T_n - T_0 = 2 \gamma / (\rho L a_n)$
and $\bar{\ell} = k_s / (\rho c_{ps} \overline{\Delta T}K)$. 
The dimensionless volumetric heat sources given by 
$q_i \bar{\ell}^2 / (\rho c_{ps} \overline{\Delta T})$
characterise the temperature rises that occur on the 
diffusive time scale due to
absorption relative to the nucleation temperature
$\overline{\Delta T} = T_n - T_0$.
These relative temperature rises are
expected to be small so the volumetric source terms are
neglected from the model, \emph{i.e.,} we take
$q_i \bar{\ell}^2 / (\rho c_{ps} \overline{\Delta T}) \simeq 0$.
The dimensionless bulk equations for the temperatures can
be written as
\begin{subequations}
  \label{app:ln}
  \begin{alignat}{2}
    \PD{\bar{\theta}_s}{\bar{\tau}} &= \nabla^2 \bar{\theta}_s, 
    &\quad \bar{\bX} \in \bar{\Omega}_s(\bar{\tau}), \\
    \hat{c}_p\PD{\bar{\theta}_l}{\bar{\tau}} &= \hat{k} \nabla^2 
    \bar{\theta}_l, &\quad \bar{\bX} \in \bar{\Omega}_l(\bar{\tau}),
  \end{alignat}
  which have initial conditions
  $\bar{\theta}_s = \bar{\theta}_l = 1$ when $\bar{\tau} = 0$
  and far-field conditions $\bar{\theta}_s \sim 1$ for
  $|\bar{\bX}| \to \infty$. At the free boundary, the
  Stefan condition reads
  \begin{align}
    \bar{v} = \bar{\beta}^{-1}\left(\PD{\theta_s}{n} - \hat{k}
      \PD{\theta_l}{n}\right), \quad \bar{\bX} \in 
    \bar{\Gamma}(\bar{\tau}),
  \end{align}
  where the Stefan number is now given by 
  $\bar{\beta} = L / (c_{ps} \overline{\Delta T})$.
  The anisotropic kinetic condition is
  \begin{align}
    \bar{v} = \bar{\theta}_I f(\psi), \quad
    \bar{\bX} \in     \bar{\Gamma}(\bar{\tau}),
  \end{align}
  The initial interface $\bar{\Gamma}(0)$ is assumed to
  be a circle of dimensionless radius
  $\bar{\alpha} = a_n / \bar{\ell}$.
\end{subequations}

In order to obtain the same asymptotic regimes as in the
early-nucleation case considered in the main text,
we let $\bar{\beta}^{-1} = \bar{b} \epsilon$ and require
the dimensionless initial melt radius to satisfy
$\bar{\alpha} \ll \epsilon$. The condition $T_n \gg T_0$
imposes an additional restriction on the dimensionless
nucleation radius given by 
$\bar{\alpha} \ll 2 \gamma / (\rho L \bar{\ell})$.
Thus, in dimensional terms, we require
\begin{align}
  a_n \ll \min\left\{\epsilon \bar{\ell},\ 
    \frac{2\gamma}{\rho L}
  \right\}.
\end{align}
We now summarise the early-time, $\bar{\tau} \ll \bar{\alpha}$,
and order-one time, $\bar{\tau} = O(1)$, problems in the
late-nucleating regime.

The early-time problem valid for $\bar{\tau} \ll \bar{\alpha}$
can be deduced by repeating the
analysis of Sec.~\ref{subsec:early}.
The lack of a volumetric heat source means that
the leading-order temperatures 
(in $\bar{\alpha}$)
are constant in time, 
$\bar{\theta}_l = \bar{\theta}_s \equiv 1$.
Thus, the anisotropic kinetic condition becomes
$\bar{v} = f(\psi)$, which is now autonomous in
the time variable $\bar{\tau}$. As a consequence, the
growth kinetics of the melt are modified.
For the anisotropy function given by
$f(\psi) = (\epsilon^2 + \sin^2 \psi)^{1/2}$,
we find that the interface can be written 
parametrically as
\begin{align}
  \bar{X} = \left[\bar{\alpha} + \frac{\bar{\tau}(1 + \epsilon^2)}
    {(\epsilon^2 + \cos^2 \varphi)^{1/2}}\right]\cos\varphi, 
  \quad
  \bar{Z} = \left[\bar{\alpha} + \frac{\bar{\tau}\epsilon^2}
    {(\epsilon^2 + \cos^2 \varphi)^{1/2}}\right]\sin\varphi, 
\end{align}
where $\varphi\in[0, 2\pi)$. Thus, the thickness and rim of
the melt now grow linearly with time rather than quadratically.
However, the morphological characteristics of the interface
remain the same as in the early-nucleation regime
and, in particular, the kinetic Wulff shapes are still
approached in the longer term. Similar changes are seen for 
other anisotropy functions as well; that is, the powers
of $\tau$ in the growth laws are reduced by a factor of
two.

For $O(1)$ times and the anisotropy function
$f = (\epsilon^2 + \sin^2 \psi)^{1/2}$, the analysis in
Sec \ref{subsec:orderone} can also be repeated in order to
derive a simplified model that collapses the melt region
onto the $\bar{Z} = 0$ axis. In particular, the 
temperature in the solid satisfies the equation
\begin{subequations}
  \label{app:O1_model}
\begin{align}
  \PD{\bar{\theta}_s}{\bar{\tau}} = \nabla^2 \bar{\theta}_s, \quad \bar{Z} > 0,
\end{align}
with $\bar{\theta}_s = 0$ when $\bar{\tau} = 0$
and $\bar{\theta}_s \sim 1$ as $|\bX| \to \infty$.
The boundary conditions on $\bar{Z} = 0$ are given by
\begin{alignat}{2}
  \PD{\bar{h}}{\bar{\tau}} &= \bar{b}\, \PD{\bar{\theta}_s}{\bar{Z}}, &\quad 
  \bar{Z} &= 0,\ \bar{R} < \bar{S}(\bar{\tau}), \\
  \PD{\bar{h}}{\bar{\tau}} &= 
  \bar{\theta}_s\Bigg[1 + \bigg(\PD{\bar{h}}{\bar{R}}
  \bigg)^2\Bigg]^{1/2}, &\quad \bar{Z} &= 0,\ \bar{R} < \bar{S}(\bar{\tau}), \\
  \PD{\bar{S}}{\bar{\tau}} &= \bar{\theta}_s, 
  &\quad 
  \bar{Z} &= 0,\ \bar{R} = \bar{S}(\bar{\tau}), \\
\PD{\bar{\theta}_s}{\bar{Z}} &= 0, &\quad \bar{Z} &= 0,\ 
  \bar{R} > \bar{S}(\bar{\tau}).
\end{alignat}
Finally, we require that
\begin{align}
  \bar{h}(\bar{S}(\bar{\tau}), 0, \bar{\tau}) = 0.
\end{align}
\end{subequations}


\end{document}